\documentclass[a4paper,11pt]{article}
\pdfoutput=1 
             
\usepackage{jheppub_mod} 

\makeatletter   
\g@addto@macro\bfseries{\boldmath}   
\makeatother   

\usepackage[utf8]{inputenc} 

\usepackage{tensor} 

\newcommand{\al}{\alpha}

\newcommand{\hal}{\frac{1}{2}}

\newcommand{\bra}[1]{\langle #1 |}
\newcommand{\ket}[1]{| #1 \rangle}

\newcommand{\del}{\partial}
\newcommand{\mac}{\mathcal}

\newcommand{\iu}{\mathrm{i}}	
\newcommand{\dd}{\mathrm{d}}	
\newcommand{\ee}{\mathrm{e}}	
\newcommand{\N}{\mathbb{N}}	
\renewcommand{\Re}{\operatorname{Re}}
\renewcommand{\Im}{\operatorname{Im}}

\newcommand{\te}[1]{\textnormal{#1}}

\graphicspath{{images/}}

\usepackage[sort&compress]{natbib} 

\usepackage{IEEEtrantools} 
\usepackage[font=small]{caption} 
\usepackage{subcaption} 
\usepackage[section]{placeins} 

\usepackage{tikz}
\usetikzlibrary{arrows,calc,tikzmark}
\tikzstyle{spring}=[thick,decorate,decoration={aspect=0.5, segment length=1.5mm, amplitude=1mm,coil}]
\usetikzlibrary{decorations.pathreplacing}
\usetikzlibrary{decorations.pathmorphing}

\title{The Casimir effect \\ in the presence of infrared transparency}

\author{Max Warkentin}

\affiliation{Arnold Sommerfeld Center for Theoretical Physics, Ludwig-Maximilians-Universit\"at, \\ Theresienstra{\ss}e 37, 80333 M\"unchen, Germany}

\emailAdd{max.warkentin@physik.uni-muenchen.de}

\abstract{We revisit the Casimir effect perceived by two surfaces in the presence of infrared (IR) transparency. To address this problem, we study a model, where such a phenomenon naturally arises: the DGP model with two parallel 3-branes, each endowed with a localized curvature term. In that model, the ultraviolet modes of the 5-dimensional graviton are suppressed on the branes, while the IR modes can penetrate them freely. First, we find that the DGP branes act as "effective" (momentum-dependent) boundary conditions for the gravitational field, so that the (gravitational) Casimir force between them emerges. Second, we discover that the presence of an IR transparency region for the discrete modes modifies the standard Casimir force --- as derived for ideal Dirichlet boundary conditions --- in two competing ways: i) The exclusion of soft modes from the discrete spectrum leads to an increase of the Casimir force. ii) The non-ideal nature of the boundary conditions gives rise to a "leakage" of hard modes. As a result of i) and ii), the Casimir force becomes weaker. Since the derivation of this result involves only the localized kinetic terms of a quantum field on parallel surfaces (with codimension one), the derived Casimir force is expected to be present in a variety of setups in arbitrary dimensions.
}

\begin{document} 
\maketitle
\flushbottom

\section{Introduction}

The purpose of this work is to investigate how the phenomenon of infrared (IR) transparency affects the Casimir force between two surfaces \cite{Casimir:1948dh}. We will study this effect by using a model, where such a phenomenon naturally arises: the DGP model with two parallel branes.\footnote{In fact, the IR transparency phenomenon originated from the DGP model \cite{dvali20004d}. Its physical meaning was then explained in Ref.~\cite{Dvali2001a}, where also the term \emph{infrared transparency} was coined. Later, Ref.~\cite{Dvali:2006su} showed that any ghost-free, large distance modified gravity theory should exhibit such a phenomenon. Furthermore, this phenomenon was generalized for massless gauge fields \cite{Dvali:2000rx}.}

The DGP model was originally introduced to study the cosmological constant problem and supersymmetry breaking \cite{dvali20004d}. However, beyond that it has turned out to be a very fruitful theoretical laboratory to study gravitational properties such as e.g.\ strong coupling in gravitational theories with more than two degrees of freedom \cite{Deffayet:2001uk}. The DGP model was the first theory of large distance modified gravity that had a ghost-free, generally-covariant, non-linear completion. It remains a rare example of a calculable, consistent effective theory that modifies spin-2 gravity at large distances while reproducing the predictions of general relativity (GR) at intermediate scales. In the present work, we will employ that model to study the gravitational Casimir effect and its modification due to the IR transparency phenomenon.

The DGP model introduces an additional, 5th spacetime dimension alongside the 4+1-dimensional (5D) version of GR that "lives" in the bulk. Moreover, it postulates a tensionless 3-brane, embedded in the 5D spacetime, which is endowed with the usual 4D Einstein-Hilbert action. The DGP action is given by
\begin{equation}
S = M_*^3 \int \dd^4 x \, \dd y \, \sqrt{|G|} \mac{R}_5 + M_{\textnormal{P}}^2  \int \dd^4 x \, \sqrt{|g|} \mac{R}_4\,, \label{eq:DGP_action}
\end{equation}
where $M_*$ is the fundamental (5D) Planck mass, while $M_{\text{P}}$ is the 4D Planck mass, observed in our world. The bulk action, where the extra (5th) dimension is labeled by the coordinate $y$, contains the bulk metric, with its determinant $|G|$ and the bulk Ricci scalar $\mac{R}_5$, whereas $|g|$ (with $g_{\mu \nu}(x^{\mu})=G_{A B} (x^{\mu},y=0)$) and $\mac{R}_4$ are the corresponding quantities on the brane.\footnote{As usual, capital roman and greek letters take values $A \in \{0,1,2,3,5 \}$ and $\mu \in \{0,1,2,3 \}$, respectively.} A possible origin for the localized curvature term on the brane is to be induced by the Standard Model fields that renormalize the graviton propagator via loop corrections.\footnote{The Standard Model fields are also localized on the 3-brane that therefore constitutes our visible universe. However, we do not display the corresponding action, since we are only interested in the effects of the gravitational part in this paper.} One of the interesting features of the theory in \eqref{eq:DGP_action} is that the graviton propagates as 4-dimensional over length scales smaller than $r_c$, but changes its behavior to 5-dimensional propagation for scales larger than $r_c$. This property is sometimes described as that the graviton is a resonance, which decays into a tower of Kaluza-Klein (KK) modes for large distances. The \emph{cross-over scale} is set by $r_c \equiv \frac{M_{\te{P}}^2}{M_*^3}$. One way to think about this behavior is to note that the brane suppresses hard modes (frequencies larger than $r_c^{-1}$) of the graviton, while it is transparent to soft modes (frequencies smaller than $r_c^{-1}$). This phenomenon is called \emph{infrared transparency}. It has been demonstrated explicitly, using a 4D mode expansion (into KK modes), for the case of a single brane in Refs.~\cite{Dvali:2001gm,Dvali2001b} (for a compact extra dimension and an infinite extra dimension, respectively).

Let us briefly review the mechanism. The mode functions $\psi_m(y)$ of the KK modes (i.e.\ the basis functions of the space along the extra dimension) have the form
\begin{equation*}
|\psi_m(y=0)| \propto \frac{1}{\sqrt{4+m^2 r_c^2}} 
\end{equation*}
on the brane ($y=0$), with $m$ the masses of the KK modes.\footnote{We will derive the mode functions explicitly for the case of two branes in section~\ref{sec:quantizing} and appendix~\ref{app:mode}.} Hence, the amplitude of the 5D field, which goes like $\sim \psi_m(y)$, is suppressed on the brane and vanishes for KK modes with $m \gg r_c^{-1}$. This affects the potential energy between two sources (separated by distance $r$), which is given by
\begin{equation*}
V(r,y=0) \propto - \frac{1}{M_*^3} \int \limits_0^{\infty} \dd m \, |\psi(y=0)|^2 \frac{\ee^{-m r}}{r}\,,
\end{equation*}
leading to a 4D gravitational potential ($1/r$) for distances $r \ll r_c$ and a 5D gravitational potential ($1/r^2$) for $r \gg r_c$.

Since a setup with a single brane and a compact extra dimension is equivalent to a setup with infinitely many branes, Ref.~\cite{Dvali:2001gm} also showed that the phenomenon of IR transparency is independent of the number of branes. We want to note, however, that the system in that paper is nonetheless not equivalent to the system, which we will consider in the present work: two parallel branes separated along an infinite extra dimension.\footnote{In fact, in the present paper, we will consider the generalized situation of two branes in a compact dimension, for reasons that will be explained in section~\ref{sec:quantizing}. At the end, however, we will send the size of the extra dimension to infinity.} This fact is illustrated in Ref.~\cite{Warkentin:2019caf}, where a modification of the gravitational laws as compared to the "ordinary" DGP model (containing just a single brane) has been discovered. For example, it has been shown that (in the limit of $R \ll r_c $, where $R$ is the separation of the branes) the gravitational potential between two sources on the same brane interpolates between the standard 4D potential $V_{\te{4D}}(r)$, for $r \ll \rho$, and $V_{\te{4D}}(r)/2$, for $\rho \ll r \ll r_c $. The length scale $\rho \equiv \sqrt{R r_c}$, which does not arise in Ref.~\cite{Dvali:2001gm},\footnote{We suspect that the reason for this is that in the setup with the compact dimension, there are infinitely many branes and infinitely many sources that contribute to the gravitational potential between "two" sources, thereby producing a different result.} is essential for understanding the decrease of the gravitational force in Ref.~\cite{Warkentin:2019caf}. Furthermore, we will show in this paper that the scale $\rho$ also plays an important role for the emerging Casimir force.

Another interesting implication of the localized term in \eqref{eq:DGP_action}, that can be traced back to the IR transparency phenomenon, is the following.\footnote{This was pointed out to us by G. Dvali in a private communication.} Since the brane "screens" the 5D gravitational force between two point sources on opposite sides of the brane (separated along the extra dimension), the situation is somewhat analogous to the so-called image problem from classical electrostatics. There, a point charge in the vicinity of a perfectly conducting plate effectively induces a mirror image charge on the opposite side of the plate. In the DGP scenario the brane effectively introduces a mirror image with a negative mass for a point mass in the brane's vicinity. Therefore, a source interacts non-trivially even with an empty brane: it gets repelled. This feature can already be viewed as an indication that the DGP model can provide a kind of "gravitational wave mirror".

In the light of the previous remarks, the starting point of our present investigation is the observation that this IR transparency phenomenon should have implications for the Casimir effect. In particular, we want to study a system that contains two parallel DGP branes and investigate the existence and particular form of the Casimir force between those branes.

Our first goal is to demonstrate that the Casimir effect can be derived even in the absence of ideal boundary conditions (such as Dirichlet boundary conditions). Instead, the DGP model comprises "effective" boundary conditions for the gravitational field.

Usually, it is questionable to consider the "gravitational Casimir effect", since boundary conditions for the gravitational field are not physical. In contrast to the electromagnetic field, which can be taken to vanish on perfect conducting plates (leading to the standard Casimir effect), gravity interacts very weakly with any material, which makes it difficult to realize a "gravitational wave mirror". However, as we intended to motivate above, the DGP model naturally provides such a setup that does not rely on ad hoc boundary conditions or speculative quantum gravity effects, as for example the proposal in Ref.~\cite{Minter:2009fx}. Naturally, if such a Casimir effect exists for gravitons, its observation would be a strong indication for the quantum origin of the gravitational field, since the Casimir energy is due to vacuum fluctuations of the underlying quantum field.

Our second goal is to show that there are deviations from the standard result (i.e.\ the Casimir force for two conducting plates) due to the fact that the branes are transparent to the IR modes. In particular, we will show that there are two contributing (and opposite) effects. On the one hand the presence of an IR region, where the soft modes do not "see" the branes, strengthens the Casimir force, because those modes are removed from the discrete spectrum. On the other hand the DGP branes provide only effective boundary conditions, so even the hard modes, which are close to $r_c^{-1}$, "leak" out of the interior of the two branes. We illustrate this by first considering a toy model where the soft modes are simply removed from the spectrum (see subsection~\ref{subsec:casimir_2dim_sharp}) and then comparing to the case where the effective nature of the boundary conditions is restored (see subsection~\ref{subsec:casimir_2dim_transition}).

This emergent Casimir force, and its deviation from the usual form, is another signature of the distinctive features of the DGP model. Hence, besides of its theoretical importance, it can lead to new ways to experimentally probe the DGP model.

Furthermore, an additional force between two branes, in the presence of localized kinetic terms, can have important implications for many braneworld scenarios as for example the brane inflation scenario \cite{dvali1999brane}, where an attractive potential between the branes drives inflation.

Finally, let us stress that this effect is generic. To derive our results, we will use a toy model that only involves a quantum scalar field with a localized kinetic term on a lower dimensional surface. As we will explain in section~\ref{sec:quantizing}, we expect that the results for bosonic fields with spins higher than 0 will not change except for numerical factors accounting for the additional degrees of freedom. Also, we will see that the qualitative result does not depend on whether we consider a 2D or a 5D system, as long as we are dealing with codimension-one branes. Therefore, such a Casimir effect would arise in a number of setups, including parallel 2-dimensional surfaces in our 4D world. In fact, the feature of IR transparency exists for many surfaces, such as ordinary walls that act as dielectrics for the electromagnetic field. They are transparent to radio waves, but suppress waves with higher frequencies (since the effective kinetic terms of the photon are different in the vacuum and inside the wall).\footnote{Note that this is reminiscent of the so-called Dvali-Shifman mechanism proposed in Ref.~\cite{dvali1997domain}, where the brane can also be considered a dielectric, while the bulk is a dual superconductor. In that case the massless gauge field can actually be exactly localized on a brane, because in that scenario the bulk is confining and hence does not allow the massless gauge field to escape there.} Hence, this effect can be probed and investigated in any experiment, where such surfaces are present.  

After the present work had been completed, it was pointed out to us by the referee that the observation that delta-function potentials lead to a modified Casimir effect has already been made in previous works. The first paper that treated delta-function potentials (often called \emph{semi-transparent potentials} in the literature) in 3+1 dimensions was Ref.~\cite{Bordag:1992cm}. Later, Ref.~\cite{Graham:2002xq} (see also Ref.~\cite{Graham:2002fw} for further discussion) included such potentials in 1+1 dimensions, and Ref.~\cite{Graham:2003ib} further clarified the situation in 3+1 dimensions (see also Refs.~\cite{Milton:2004vy,Milton:2004ya} for additional results and an elucidating discussion concerning delta-function potentials in both 1+1 and 3+1 dimensions). However, in those papers delta-function potentials are coupled to the fields, rather than to their derivatives. Consequently, the resulting Casimir force, although different from the standard result (in the case of Dirichlet boundary conditions), is not equal to the one derived in the present work. Although those delta-function potentials have been studied extensively since then, to the best of our knowledge, the particular setup with derivative couplings addressed in this work (that is crucial for the aforementioned IR transparency phenomenon) has not yet been studied in the literature.

Our paper is organized as follows. First, in section~\ref{sec:quantizing}, we describe the system and quantize it, showing how a discrete spectrum of modes arises. Then, in section~\ref{sec:energy}, we explain how to regularize and renormalize the vacuum energy within the dimensionally reduced description. In section~\ref{sec:casimir_2dim} we derive the Casimir effect in 1+1 dimensions, since some of the analytical expressions can only be obtained in this simplified system. There, in the first two subsections, \ref{subsec:casimir_2dim_sharp} and \ref{subsec:casimir_2dim_transition}, we discuss the effect of the IR transparency region on the Casimir force and explain how the effective nature of the boundary conditions provided by the DGP branes further modifies the final result. This is done by deriving approximate analytical expressions. The result is then further justified and refined by a numerical analysis in the subsection~\ref{subsec:casimir_2dim_numerics}. In section~\ref{sec:casimir_5dim} we finally treat the 5D system, obtaining most of the results numerically. We then conclude in section~\ref{sec:conclusion}.

The actual calculation of the mode functions, necessary for a dimensional reduction, and derivation of the mass quantization equation is performed in appendix~\ref{app:mode}. Appendix~\ref{app:sum} contains the evaluation of a Bessel function sum, necessary for the analytical calculation of the 5D Casimir energy in the limit $r_c \to \infty$.

\section{Dimensional reduction and quantization} \label{sec:quantizing}

Let us consider a theory with the action
\begin{equation}
S = \int \mbox d^4 x \, \dd y \, \left\{ \hal \left( \del_A \Phi \right)^2 +  r_c \left[ \delta \left(y + \frac{R}{2} \right) + \delta \left(y-\frac{R}{2} \right) \right] \hal \left( \del_{\mu} \Phi \right)^2 \right\}\,, \label{eq:action}
\end{equation}
where $y \in (-\infty,\infty)$. This theory is a simplified version of \eqref{eq:DGP_action}, but it captures the essential features of the DGP model, necessary for our discussion, because of the following reasons. First, we only need to consider the linearized version of the DGP action, because we are interested in the vacuum energy of the quantum field. Second, although the theory \eqref{eq:action} describes one propagating (scalar) degree of freedom, whereas \eqref{eq:DGP_action} describes 5 propagating degrees of freedom (scalar, vector and tensor),\footnote{Note that a modified gravitational theory such as \eqref{eq:DGP_action} exhibits strong coupling \cite{Deffayet:2001uk}. However, since the Casimir effect is an IR effect, which persists after cutting off the high energy behavior of the system, this should not modify our result.} the standard Casimir effect is qualitatively the same for all bosonic fields (see e.g.\ Ref.~\cite{bordag2009advances} for spin-0 and spin-1 and Refs.~\cite{Lin:2000ef,Quach_2015} for spin-2). The different numbers of degrees of freedom just affect the numerical factor of the result. As we will show, the modification of our derived Casimir effect is entirely due to the special mass quantization resulting from the presence of the branes. Hence, we do not expect differences for higher spin fields, besides the usual factors for the degrees of freedom.

For our purposes we will need an IR regulator, so we introduce the space as finite with size $L$ in the extra dimension and size $V$ in the transverse dimensions (on the brane).\footnote{The compactification of the 3-volume on the brane is not necessary, but we will keep it until the next section for clarity of presentation.} We choose periodic boundary conditions for our space. Later, we will remove the regulator and end up with an infinite space (in all dimensions).

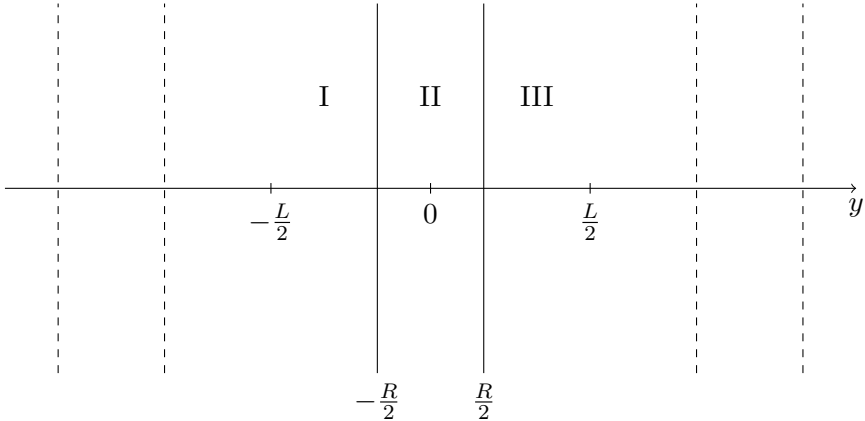
\begin{figure}[t]
\centering
\begin{tikzpicture}[scale=0.70]

             
                  \draw[->] (-8,0) -- (8,0) node[anchor=north] {$y$} ;
      \draw (0,0.1) -- (0,-0.1) node[anchor=north] {0} ;
        \draw (-3,0.1) -- (-3,-0.1) node[anchor=north] {$-\frac{L}{2}$} ;
             \draw (3,0.1) -- (3,-0.1) node[anchor=north] {$\frac{L}{2}$} ;
      
      \draw (-1,3.5) -- (-1,-3.5) node[anchor=north] {$-\frac{R}{2}$} ;
      \draw (1,3.5) -- (1,-3.5) node[anchor=north] {$\frac{R}{2}$} ;
      
%

  \draw[dashed] (-7,-3.5) -- (-7,3.5) ;
               \draw[dashed] (-5,-3.5) -- (-5,3.5) ;
               
                     \draw[dashed] (7,-3.5) -- (7,3.5) ;
               \draw[dashed] (5,-3.5) -- (5,3.5) ;
               
               \draw (-2,1.75) node[anchor=center] {I};
                     \draw (0,1.75) node[anchor=center] {II};
          \draw (2,1.75) node[anchor=center] {III};

\end{tikzpicture}
\caption{The setup: Periodic space (with period $L$) along the extra dimension $y$, with two DGP branes (solid lines). The branes (with suppressed transverse dimensions) are repeated (dashed lines) after the period. I, II and III are the regions we need to match in order to find the field solution.}
\label{fig:setup}
\end{figure}

Focussing on the extra dimension and suppressing the transverse dimensions, the system is illustrated by Figure~\ref{fig:setup}.

We can decompose the 5D scalar field as
\begin{equation*}
\Phi(x^{\mu},y)= \sum \limits_{\alpha=1}^2 \sum \limits_m  \psi_{m,\alpha}(y) \phi_{m,\alpha}(x^{\mu})\,,
\end{equation*}
where the mode functions $\psi_{m,\alpha}(y)$ span a complete basis of the $y$-space, satisfying the equation
\begin{equation}
\left\{ \del_y^2 + m_{\al}^2 \left[ 1+r_c \, \delta \left(y + \frac{R}{2}\right) + r_c \, \delta \left(y-\frac{R}{2} \right) \right] \right\} \psi_{m,\alpha}(y) =0\,, \label{eq:schroedinger}
\end{equation}
subject to the periodic boundary condition. Since the system is symmetric around the origin, we have divided the mode functions into even ($\al=1$) and odd ($\al=2$). We derive such mode functions in appendix \ref{app:mode}. Using that they satisfy the orthonormality condition
\begin{equation}
\int \limits_{- \frac{L}{2}}^{\frac{L}{2}} \dd y \, \psi_{m,\alpha}(y) \psi_{m',\alpha'}(y) \left[ 1+r_c \, \delta \left(y + \frac{R}{2}\right) + r_c \, \delta \left(y-\frac{R}{2} \right) \right] = \delta_{m,m'} \delta_{\alpha, \alpha'}\,, \label{eq:completeness}
\end{equation}
the 5D action reduces to
\begin{equation}
S=\int \dd^4 x \, \sum_{\alpha} \sum \limits_ m \hal \left[ \left( \del_{\mu} \phi_{m,\alpha}(x) \right)^2 - m_{\al}^2 \phi^2_{m,\alpha}(x) \right]\,, \label{eq:KK_action} 
\end{equation}
where we use the shorthand notation $x \equiv x^{\mu}$. The information about the extra dimension and the brane configuration is now encoded in the discrete masses. By solving for $\psi_{m,\alpha}(y)$, we found (see appendix~\ref{app:mode}) that the masses have to satisfy the following quantization equation:
\begin{IEEEeqnarray}{rCl} \label{eq:masses}
\tan \frac{m_\al L}{2} &=& - \frac{m_\al r_c}{2} \times \left\{ 
\begin{aligned} 
&\frac{1+\cos m_\al R}{1+\frac{m_\al r_c}{2} \sin m_\al R}\,, &  \mbox{for } \alpha=1 \te{ (even)}& ,  \\ 
&\frac{1-\cos m_\al R}{1-\frac{m_\al r_c}{2} \sin m_\al R}\,, &  \mbox{for } \alpha=2 \te{ (odd)}& .
\end{aligned} \right.  
\end{IEEEeqnarray} 
Note that we only have one mode with zero mass, the even zero-mode, since the odd zero-mode vanishes, $\psi_{m=0,\alpha=2}(y)=0$. 

Thus, we traded the extra dimension for a KK tower of massive Klein-Gordon fields satisfying the equation
\begin{equation}
\left( \Box + m_{\al}^2 \right) \phi_{m,\alpha}(x) = 0\,. \label{eq:4dim_KG}
\end{equation}
Using the finite 3-volume on the brane, with size $V$ and periodic boundary conditions, the solutions to \eqref{eq:4dim_KG} can be quantized in the standard way, leading to
\begin{equation*}
\widehat{\phi}_{m,\alpha}= \frac{1}{\sqrt{V}} \sum \limits_{\vec{k}} \frac{1}{\sqrt{2 \omega_{m,\al}(\vec{k})}} \left( \widehat{a}_{m,\al}(\vec{k}) \ee^{- \iu \left( \omega_{m,\al}(\vec{k}) t - \vec{k} \vec{x} \right) } + \widehat{a}^\dagger_{m,\al}(\vec{k}) \ee^{ \iu \left( \omega_{m,\al}(\vec{k}) t - \vec{k} \vec{x} \right) } \right)\,,
\end{equation*}
with 
\begin{equation*}
\omega_{m,\al}(\vec{k})=\sqrt{|\vec{k}|^2 + m_{\al}^2} , \qquad k_i = \frac{2 \pi n_i}{V^{1/3}}, \quad n_i=0,\pm1,\ldots
\end{equation*}
and the canonical commutation relations
\begin{equation*}
\left[ \widehat{a}_{m,\al}(\vec{k}) , \widehat{a}^\dagger_{m',\al'}(\vec{k'}) \right] = \delta_{\vec{k},\vec{k'}} \delta_{m,m'} \delta_{\alpha, \alpha'} , \qquad \te{(all others zero)}\,.
\end{equation*}
The Hamiltonian is then given by
\begin{equation*}
\widehat{H} = \sum_{\alpha} \sum \limits_ m \sum \limits_{\vec{k}} \omega_{m,\al}(\vec{k}) \left( \widehat{a}^{\dagger}_{m,\al}(\vec{k}) \widehat{a}_{m,\al}(\vec{k}) + \hal \right)\,.
\end{equation*}

\section{Vacuum energy} \label{sec:energy}

The vacuum energy per unit 3-volume (i.e.\ the vacuum energy density measured by a brane-observer) is
\begin{equation*}
E = \frac{1}{V} \bra{0} \widehat{H} \ket{0} = \frac{1}{V} \sum_{\alpha} \sum \limits_ m \sum \limits_{\vec{k}} \frac{\omega_{m,\al}(\vec{k})}{2}\,.
\end{equation*}
Let us now go to the limit of an infinitely large 3-brane ($V \to \infty$). Then we can replace
\begin{equation*}
\frac{1}{V} \sum \limits_{\vec{k}} \leftrightarrow \int \frac{\dd k_1 \, \dd k_2 \, \dd k_3}{(2 \pi)^3}
\end{equation*}
and find
\begin{equation*}
E = \frac{1}{4 \pi^2} \sum_{\alpha} \sum \limits_ m \int \limits_0^{\infty} \dd k \, k^2 \omega_{m,\al}(\vec{k})\,,
\end{equation*}
where we used spherical coordinates with $k \equiv |\vec{k}|$. Of course, this quantity is divergent, because we are integrating over all momenta and summing over the full KK tower of masses. In order to regularize this expression, we will use the scheme of cut-off regularization (there is a vast amount of literature on different regularization schemes, see e.g.\ Ref.~\cite{bordag2009advances}) and introduce the exponential cut-off
\begin{equation*}
\exp\left[ -\omega_{m,\al}(\vec{k}) \frac{a}{\pi} \right]\,.
\end{equation*}
Thus, we are cutting off the high frequencies at the frequency $\sim 1/a$. We can interpret this in an analogous way to the standard calculation of the vacuum energy between two conducting plates in the presence of an electromagnetic field. There, the plates are not expected to provide Dirichlet boundary conditions for the photon at frequencies much larger than the plasma frequency of the plates. Here, we also expect the effective Lagrangian \eqref{eq:action} to be valid only for distances larger than $a$. Therefore, the regularized vacuum energy is
\begin{equation}
E^{\te{reg}}  = \frac{1}{4 \pi^2} \sum_{\alpha} \sum \limits_ m \int \limits_0^{\infty} \dd k \, k^2 \omega_{m,\al}(\vec{k}) \ee^{-\omega_{m,\al}(\vec{k}) \frac{a}{\pi}}\,. \label{eq:E_reg_5dim}
\end{equation}

By construction, expression \eqref{eq:E_reg_5dim} is cut-off dependent and will blow up for $a \to 0$. However, since this singular behavior has nothing to do with the presence of the branes, but comes from the fact that the vacuum energy is divergent, we will renormalize the energy by subtracting the vacuum energy of a system without branes. Note, however, that such a subtraction does not remove the divergence completely. We will find in section~\ref{subsec:casimir_2dim_numerics}, using numerical methods, that the renormalized energy in 2D still has a non-vanishing, logarithmic $a$-dependence. Also, in the 5D system an $a$-dependence remains. However, this divergent part is independent of the separation distance of the branes. Hence, the Casimir force is finite and cut-off independent.\footnote{Note that there exists extensive literature that deals with improved regularization and renormalization techniques such that those kinds of remaining divergences are removed. In the context of delta-function potentials one can also find discussions of the physical origin of such divergences (see e.g.\ Refs.~\cite{Graham:2003ib,Milton:2004ya}). Since we are only interested in the observable Casimir force, it is beyond the scope of this paper to discuss this further.}

Now, in the system without the branes, we can again decompose the 5D scalar field into massive KK modes, using just $\cos{m y}$ and $\sin{m y}$ as mode functions. The periodic boundary conditions will then lead to
\begin{equation}
m_0= \frac{2 \pi n}{L} , \quad n=0,1,2,\ldots \label{eq:plane_masses}
\end{equation}
Note that here, unlike in the setup with branes, the masses of the even and the odd modes are equal (and again only the even modes have a zero-mode, because $\sin(0)=0$).\footnote{The reason that in the dimensionally reduced system every mode (except the even zero-mode) is twofold degenerate, is that the original 5D setup (without branes) had a conserved 5th momentum due to the translational invariance in $y$-direction, which translates to a U(1)-invariance in the reduced system.} Further notice that the masses $m_0$ can also be recovered from \eqref{eq:masses} in the limit $r_c \to 0$, as it should be.

Thus, the (regularized) vacuum energy in the absence of branes is given by
\begin{equation*}
E_0^{\te{reg}}  = E^{\te{reg}}|_{m \to m_0}\,.
\end{equation*}
Then, the Casimir energy of the DGP system is
\begin{equation}
E_{\te{C}} \equiv  \lim_{\substack{a \to 0 \\ L \to \infty}} \left( E^{\te{reg}}  - E_0^{\te{reg}} \right)\,, \label{eq:casimir_energy}
\end{equation}
where we removed the short distance cut-off $a$ and sent the size of the extra dimension to infinity (which is usually the case in the DGP model). We expect expression \eqref{eq:casimir_energy} to be finite, since the Casimir effect is known to be an IR effect and therefore should be independent of high energy physics. We will now calculate this expression and show that this is indeed the case (note, however, the comment beneath \eqref{eq:E_reg_5dim}).

\section{Casimir effect in 1+1 dimensions} \label{sec:casimir_2dim}

Since evaluating expression \eqref{eq:E_reg_5dim} is quite involved, we will first consider a simplified version in order to isolate the relevant features, before tackling the full problem.

Let us consider our setup in 1+1 dimensions, i.e.\ with just the "extra" dimension and time. In this case, the 3-branes become 0-branes (i.e.\ just points in the $y$-direction) and there are no transverse dimensions on the branes. This situation is described by the action
\begin{equation}
S_{\te{2D}} = \int \mbox d t \, \dd y \, \hal \left\{ \dot{\Phi}^2 - \left( \del_y \Phi \right)^2 +  r_c \left[ \delta \left(y + \frac{R}{2} \right) + \delta \left(y-\frac{R}{2} \right) \right]  \dot{\Phi}^2 \right\}\,,
\end{equation}
where $(\, \dot{} \,)$ denotes time-derivatives.

Now we decompose
\begin{equation*}
\Phi(t,y)= \sum \limits_{\alpha=1}^2 \sum \limits_m  \psi_{m,\alpha}(y) \phi_{m,\alpha}(t)\,,
\end{equation*}
where $\psi_{m,\alpha}(y)$ are the same mode functions as before, satisfying \eqref{eq:schroedinger}. Then, the dimensionally reduced action reads
\begin{equation*}
S_{\te{2D}}=\int \dd t \, \sum_{\alpha} \sum \limits_ m \hal \left( \dot{\phi}^2_{m,\alpha}(t) - m_{\al}^2 \phi^2_{m,\alpha}(t) \right)\,. 
\end{equation*}
The KK fields are now just harmonic oscillators with the frequencies given by the solutions of \eqref{eq:masses}. Following the same steps as in section~\ref{sec:quantizing}, we find the Hamiltonian
\begin{equation*}
{}^{(\te{2D})}\widehat{H} = \sum_{\alpha=1}^2 \sum \limits_ {m>0} m_\al \left( \widehat{a}^\dagger_{m,\al} \widehat{a}_{m,\al} + \hal \right)\,.
\end{equation*}
Note that in the 2D case the Hamiltonian does not contain the zero-mode (with $m_1=0$), because it is constant and hence drops out ($\dot{\phi}_{m=0,\al=1}=0$).

The vacuum energy is given by
\begin{equation*}
\tensor[^{(2\te{D})}]{E}{} = \bra{0} {}^{(2\te{D})}\widehat{H} \ket{0} = \sum_{\alpha=1}^2 \sum \limits_ {m>0} \frac{m_\al}{2}\,,
\end{equation*}
and after regularizing we get
\begin{equation}
\tensor[^{(2\te{D})}]{E}{^{\te{reg}}}  = \sum_{\alpha=1}^2 \sum \limits_ {m>0}  \frac{m_\al}{2} \ee^{-m_\al \frac{a}{\pi}}\,,  \label{eq:E_reg_2dim}
\end{equation}
which is much simpler than \eqref{eq:E_reg_5dim}. However, since the masses $m_\al$ cannot be determined exactly, we can calculate \eqref{eq:E_reg_2dim} only either approximately or numerically. We will do the former in the next two sections and the latter in the section after that.

\subsection{IR transparent and opaque regions --- sharp transition approximation} \label{subsec:casimir_2dim_sharp}

In this paper we are interested in the parameter space covering $R \lesssim r_c$.\footnote{The same setup (with $R \lesssim r_c$) has also been studied in Ref.~\cite{Warkentin:2019caf}. However, the system with $R \gtrsim r_c$ might also be interesting to investigate in the future.} If we particularly consider the limit $R \ll r_c$, we find that \eqref{eq:masses} allows two different regimes. We can rewrite it as
\begin{equation} 
- \frac{2 \sin \frac{m_\al L}{2}}{m_\al r_c} = \left\{ 
\begin{aligned} 
&2 \cos \left[ \frac{m_\al}{2} (L-R) \right] \cos \frac{m_\al R}{2}\,, & \mbox{for } \alpha=1\,   \te{ (even)},&  \\ 
&-2 \sin \left[ \frac{m_\al}{2} (L-R) \right] \sin \frac{m_\al R}{2}\,, &  \mbox{for } \alpha=2\,  \te{ (odd)}.&
\end{aligned} \right.    \label{eq:masses_rewritten}
\end{equation}

For $m_\al r_c \ll 1$, the left-hand side of \eqref{eq:masses_rewritten} blows up and hence we find solutions for (since the right-hand side lies in the interval $[-2,2]$)
\begin{IEEEeqnarray} {rCl} \IEEEyesnumber \IEEEyessubnumber* \label{eq:infra_trans_masses}
m_{\alpha} =  \frac{2 \pi n}{L} , \quad n=1,2,3,\dots \qquad \left(m_\al \ll r_c^{-1} \right)\,.
\end{IEEEeqnarray} 
For $m_\al r_c \gg 1$, the left-hand side of \eqref{eq:masses_rewritten} vanishes, which leads to
\begin{IEEEeqnarray} {rCl} \IEEEyessubnumber* \label{eq:opaque_masses}
\begin{aligned} 
m_1 &= \left\{ 
\begin{aligned} 
&\frac{\pi n}{L-R}   \\
&\frac{\pi n}{R}   
\end{aligned} \right.\,, \quad n=1,3,5,\dots  \qquad \te{(even modes)}\,, \\[0.5ex]
m_2 &= \left\{ 
\begin{aligned} 
&\frac{\pi n}{L-R}   \\
&\frac{\pi n}{R}   
\end{aligned} \right.\,, \quad n=2,4,6,\dots  \qquad \te{(odd modes)}\,,
\end{aligned} \qquad \left(m_\al \gg r_c^{-1} \right) .
\end{IEEEeqnarray} 
Note, however, that the upper solutions of the even and odd parts of \eqref{eq:opaque_masses} are only valid for $n \gg \frac{L-R}{r_c}$. 

\begin{figure}[t]
\centering
\begin{tikzpicture}[scale=0.70]

 
 \node[circle,draw,minimum size=3.5cm] (a) at (0,0) {};

\draw (a.-60) -- ++ (-60:0.3);
\draw (a.-60) -- ++ (180-60:0.3);
\draw (a.-120) -- ++ (-120:0.3);
\draw (a.-120) -- ++ (180-120:0.3);

         \draw (a.90) node[anchor=south] {$L-R$};
             \draw (a.-90) node[anchor=north] {$R$};

%



\end{tikzpicture}
\caption{A ring (with circumference $L$) with two branes (perpendicular lines) separated by $R$. For "transparent" branes one expects the frequencies to obey $m=  \frac{2 \pi n}{L}$, with $n \in \N$. For Dirichlet boundary conditions at the branes, one expects $m =  \frac{\pi n}{L-R}$ and $m =  \frac{\pi n}{R}$, with $n \in \N$. Both solutions are the limiting cases of \eqref{eq:masses} for $m \ll r_c^{-1}$ and $m \gg r_c^{-1}$, respectively.}
\label{fig:ring}
\end{figure}
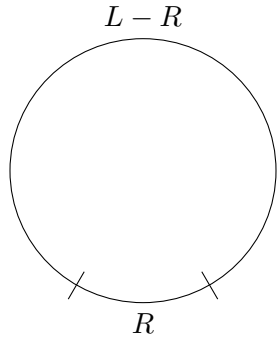

We see that the system has an \emph{infrared transparent regime} (for $m \ll r_c^{-1}$), where the modes do not "see" the branes and hence have the same frequency as in a (periodic) box of size $L$. The system also has an \emph{opaque regime} (for $m \gg r_c^{-1}$), where the branes act effectively as Dirichlet boundary conditions for the modes and hence the latter acquire frequencies as expected for such a configuration. The system is depicted in Fig.~\ref{fig:ring}.\footnote{This is of course just a slightly different way to view Fig.~\ref{fig:setup}.}

Now, let us first ignore the fact that there is a transition region at $m \sim r_c^{-1}$ and consider a "toy model", where we artificially construct boundary conditions for the scalar field in such a way that the modes with $m \leq r_c^{-1}$ are "free", i.e.\ they do not have to fulfill any boundary conditions at the location of the branes, while all heavier modes, $m > r_c^{-1}$, have to satisfy Dirichlet boundary conditions there (so the branes act as "perfect conductors" for the scalar field\footnote{We can see explicitly that the field amplitudes vanish at $y=\pm \frac{R}{2}$ for $m r_c \gg 1$ from the expressions for the mode functions \eqref{eq:modes_fct_even} and \eqref{eq:modes_fct_odd}.}). Then, the (regularized) energy for a system with such a \emph{sharp} transition is given by
\begin{equation}
\tensor*[^{(2\te{D})}]{E}{^{\te{reg}}_{\te{s}}}  =  \underbrace{\sum \limits_ {n=1}^{n_{\te{max}}}  \frac{2 \pi n}{L} \ee^{-\frac{2 a n}{L}}}_{\tensor*[]{\Sigma}{^{\te{s}}_1}} + \underbrace{\hal \sum \limits_ {n= n_{\te{min}} +1}^{\infty}  \frac{\pi n}{L-R} \ee^{-\frac{a n}{L-R}}}_{\tensor*[]{\Sigma}{^{\te{s}}_2}} +  \underbrace{\hal \sum \limits_ {n= 1}^{\infty}  \frac{\pi n}{R} \ee^{-\frac{a n}{R}}}_{\tensor*[]{\Sigma}{^{\te{s}}_3}}\,,  \label{eq:E_reg_sharp_2dim}
\end{equation}
with
\begin{equation}
n_{\te{max}}= \frac{1}{2 \pi} \frac{L}{r_c}, \qquad  n_{\te{min}}=\frac{1}{\pi} \frac{L-R}{r_c}\,. \label{eq:n_max_min}
\end{equation}

If we perform the summations and then expand around $a=0$ and $\frac{1}{L}=0$, we find
\begin{IEEEeqnarray} {rCl}  \IEEEyesnumber \IEEEyessubnumber* \label{eq:sigma_sharp_1}
\tensor*[]{\Sigma}{^{\te{s}}_1} &=& \frac{L}{4 \pi r_c^2} + \frac{1}{2 r_c} - \frac{L a}{6 \pi^2 r_c^3} + \mac{O}\left(\frac{a}{r_c^2} \right) + \mac{O}\left(\frac{a}{L r_c} \right)\,, \\
\tensor*[]{\Sigma}{^{\te{s}}_2} &=& \frac{\pi L}{2 a^2} - \frac{\pi R}{2 a^2} - \frac{L}{4 \pi r_c^2} + \frac{R}{4 \pi r_c^2} - \frac{1}{4 r_c} + \frac{L a}{6 \pi^2 r_c^3}  + \mac{O}\left(\frac{1}{L} \right) + \mac{O}\left(\frac{a}{r_c^2} \right)\,, \IEEEyessubnumber* \label{eq:sigma_sharp_2} \\
\tensor*[]{\Sigma}{^{\te{s}}_3} &=&  \frac{\pi R}{2 a^2} - \frac{\pi}{24 R} + \mac{O}\left(\frac{a^2}{R^3} \right)\,. \IEEEyessubnumber* \label{eq:sigma_sharp_3}
\end{IEEEeqnarray} 

To renormalize expression \eqref{eq:E_reg_sharp_2dim}, we again subtract the vacuum energy in the absence of branes,
\begin{equation*}
\tensor*[^{(2\te{D})}]{E}{^{\te{reg}}_0}  = \sum_ {n=1}^{\infty} \frac{2 \pi n}{L} \ee^{-\frac{2 a n}{L}} = \frac{\pi L}{2 a^2} + \mac{O}\left(\frac{1}{L} \right) + \mac{O}\left(\frac{a^2}{L^3} \right)\,.
\end{equation*}
Then, it follows that 
\begin{IEEEeqnarray}{rCl} 
\tensor*[^{(2\te{D})}]{E}{^{\te{s}}_{\te{C}}} &=&  \lim_{\substack{a \to 0 \\ L \to \infty}} \left( \tensor*[^{(2\te{D})}]{E}{^{\te{reg}}_{\te{s}}}   - \tensor*[^{(2\te{D})}]{E}{^{\te{reg}}_0}  \right) , \nonumber \\
&=& - \frac{\pi}{24 R} + \frac{1}{4 r_c} + \frac{R}{4 \pi r_c^2}\,. \label{eq:casimir_energy_sharp_2dim}
\end{IEEEeqnarray}

Thus, we find that, after removing the regulators $a$ and $L$, the resulting Casimir energy has three finite and cut-off independent terms. The first term is the same as one would get, if calculating the standard Casimir energy between two conducting plates (Dirichlet boundary conditions) separated by a distance $R$ in 1+1 dimensions.\footnote{Throughout this paper we are using units where $c=\hbar=1$.} However, we see that the inclusion of an IR transparency region leads to two new contributions that modify the Casimir energy. The first new contribution (2nd term in \eqref{eq:casimir_energy_sharp_2dim}) increases the Casimir energy by a constant. However, note that the sign of this term depends on whether the first mode in the opaque region is even or odd. In the present situation the constant is positive, because the first mode entering $\tensor*[]{\Sigma}{^{\te{s}}_2}$ in \eqref{eq:E_reg_sharp_2dim} is odd. Had we organized the system according to the prescription --- modes with $m < r_c^{-1}$ are "free" and modes with $m \geq r_c^{-1}$ are "bound" --- the first mode to enter the summation would have been the even one. In this case the constant contribution would be negative, $- \frac{1}{4 r_c}$.

Since we are ultimately interested in the Casimir force, which is given by
\begin{equation*}
F_{\te{C}} \equiv - \frac{\del E_{\te{C}}}{\del R}\,,
\end{equation*}
the more important, novel contribution is the last term in \eqref{eq:casimir_energy_sharp_2dim}. Here, the sign turns out to be independent of the particular separation of the free and bound modes (i.e.\ independent of whether the first mode is even or odd) and is always positive. Therefore, the Casimir force in the presence of an IR transparency region is
\begin{equation}
\tensor*[^{(2\te{D})}]{F}{^{\te{s}}_{\te{C}}} = - \frac{\del}{\del R} \, \tensor*[^{(2\te{D})}]{E}{^{\te{s}}_{\te{C}}} = - \frac{\pi}{24 R^2} -\frac{1}{4 \pi r_c^2}\,. \label{eq:casimir_force_sharp_2dim}
\end{equation}
Hence, it is amplified as compared to the standard Casimir force without an IR transparency region.

This result can be interpreted in the following way: Usually (so in the case where the branes provide ideal Dirichlet boundary conditions), we can "fit" only those wavelengths $\lambda$ into the system shown in Fig.~\ref{fig:ring}, which are fractions of $2 R$ and $2 (L-R)$, hence
\begin{equation}
\lambda =  \left\{ 
\begin{aligned} 
&2 \frac{R}{n}, & \qquad (\te{inside the branes}),& \\
&2 \frac{L-R}{n}, & \qquad (\te{outside the branes})\,,& 
\end{aligned} \right. \label{eq:wavelengths}
\end{equation}
where $n \in \N$. However, we have to exclude wavelengths which lie below the validity distance of the theory by introducing the cut-off $a$, hence excluding wavelengths $\lambda \lesssim a$. One then finds in the limit $L \to \infty$ (which turns the "outside"-modes continuous) and $a \to 0$ that due to the fact that the "inside"-modes are discrete (and hence reduced in number) there is an attractive force between the plates (branes). The fact that this effect is independent of the short distance cut-off $a$ is said to indicate that the Casimir effect is an IR effect. 

Now we found the following modification to this situation: If we remove all the ("outside") wavelengths from \eqref{eq:wavelengths}, for which $n \lesssim 2 \frac{L-R}{r_c}$, such that all wavelengths $\lambda \gtrsim r_c$ do not "see" the boundary conditions at the branes anymore\footnote{So the wavelengths with $\lambda \gtrsim r_c$ are given by $\lambda=\frac{L}{n}$. In the continuous limit $L \to \infty$ these wavelengths are not constrained at all.}, we diminish the number of allowed modes in our brane-system even further. Hence, the (magnitude of the) Casimir force is increased by a constant (proportional to the size of the "exclusion window").

\begin{figure}[t]
\centering
\begin{tikzpicture}

      \draw[->] (-6,0) -- (6.5,0) ;
      \draw (-3,0.1) -- (-3,-0.1) node[anchor=north] {$a$} ;
      \draw (0,0.1) -- (0,-0.1)  node[anchor=north] {$R$};  
          \draw (3,0.1) -- (3,-0.1)  node[anchor=north] {$r_c$};  
              \draw (6,0.1) -- (6,-0.1)  node[anchor=north] {$L$};     
              
               \draw (-4.5,0)  node[text width=3cm,align=center,anchor=south] {short distance physics} ;
        \draw (0,0.1) node[text width=6cm,align=center,anchor=south] {modes can be fit only inside or outside the branes}; 
         \draw (4.5,0) node[text width=3cm,align=center,anchor=south] {modes of periodic box do not "see" the branes};

\end{tikzpicture}
\caption{Different scales of the system and their relevance.}
\label{fig:scales}
\end{figure}
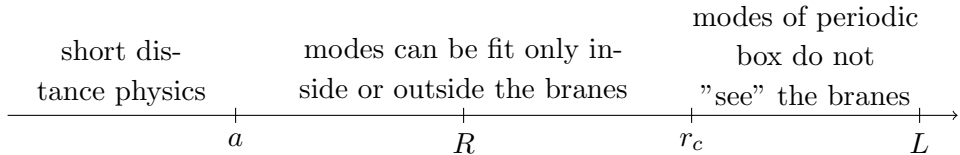

We summarize the relevant distance scales in Fig.~\ref{fig:scales}.

\subsection{Taking into account the transition region --- leaking branes} \label{subsec:casimir_2dim_transition}

In order to analyze our more realistic model, where the branes do not just alternate between being perfect conductors (for $m > r_c^{-1}$) and being invisible (for $m < r_c^{-1}$), but have the DGP-like behavior, we have to study more closely the region $m \sim r_c^{-1}$. In this case, the right-hand side of \eqref{eq:masses_rewritten} is an oscillating function and hence the solutions for $m_\al$ are not available exactly. However, working in the limit $R \ll r_c<L$, we can derive leading corrections to the results \eqref{eq:infra_trans_masses} and \eqref{eq:opaque_masses}, which improve the accurateness of the result, as we approach $m \sim r_c^{-1}$. In this limit, the "inside"-modes, $m \propto \frac{n}{R}$, of the regime $m r_c \gg 1$ are already good approximate solutions, so we can focus on the "outside"-modes, $m \propto \frac{n}{L-R}$, there.

\paragraph{Even modes}

For the correction of the even modes in the opaque region, we can expand \eqref{eq:masses_rewritten} using $m (L-R) = \pi n + \epsilon$ (for $|\epsilon| \ll 1$) and find $\epsilon \sim 2 \frac{L-R}{\pi n r_c}$, as long as $n \lesssim \frac{2 L}{\pi R}$. 

Similarly, for the IR region, we expand using $m L =2 \pi n + \epsilon$ (for $|\epsilon| \ll 1$) and find $\epsilon \sim - 4 \pi n \frac{r_c}{L}$.

Thus, in the leading approximation we obtain
\begin{IEEEeqnarray} {rCl}  \label{eq:corrected_masses_even}
m_1 = \left\{ 
\begin{aligned} 
&\frac{2 \pi n}{L} - \frac{4 \pi n}{L} \frac{r_c}{L} , \qquad  \te{for } n \ll n_{\te{max}}\,, &  \te{ (for } n \in \N),&    \\
&\left. \begin{aligned} 
&\underbrace{\frac{\pi n}{L-R} + \frac{2}{\pi n} \frac{1}{r_c}}_{m_1^{\te{c}}}\,, \qquad \te{for }   n_{\te{min}} \ll n \ll n_*\,,   \\
&\frac{\pi n}{R}\,, 
\end{aligned}  \right\} &  (\te{for odd integers }n)\,,& 
 \end{aligned} \right. 
\end{IEEEeqnarray} 
with $n_{\te{max}}$ and $n_{\te{min}}$ again given by \eqref{eq:n_max_min}. Note that the form of the correction of $m_1^{\te{c}}$ is only valid up to $n \sim n_*$, with
\begin{equation}
n_* \equiv \frac{2}{\pi} \frac{L-R}{R}\,. \label{eq:n_star}
\end{equation}

\paragraph{Odd modes}

The leading correction of the odd modes in the opaque region has a more subtle form than for the even modes. It turns out that it first grows (with $n$) and then decreases again, with the turning point set by a new scale. We can see this as follows. We rewrite \eqref{eq:masses_rewritten} using $m (L-R) = \pi n + \beta$ (for $|\beta| \ll \pi n$\footnote{Note that this time we do not assume $\beta$ to be much smaller than 1. However, we still require it to be much smaller than $\pi n$, since we are looking for a correction to the leading term of $m_2$.}) and find $\beta \sim 2 \arctan f(n)$, as long as $n \lesssim \frac{2 L}{\pi R}$. The argument of the $\arctan$ is given by
\begin{equation*}
f(n) = \frac{R}{2 L} \frac{n \pi}{ \left( \frac{n}{n_\rho} \right)^2 -1}\,,
\end{equation*}
where 
\begin{equation*}
n_\rho=\sqrt{2} \frac{L-R}{\pi \rho} \qquad \te{and} \qquad \rho=\sqrt{R r_c}\,.
\end{equation*}
Notice that first (for $n \lesssim n_\rho$) the correction grows linearly with $n$ ($\beta \propto -n \frac{R}{L}$) and later (for $n \gtrsim n_\rho$) decreases again ($\beta \propto \frac{1}{n} \frac{L}{r_c}$). At the turning point, at $n \sim n_\rho$, the mass (leading term) is $m \sim \rho^{-1}$, thus we have found a new characteristic scale, $\rho$.

Finally, for the IR region, we expand using $m L =2 \pi n + \epsilon$ (for $\epsilon \ll 1$) and find $\epsilon \sim - 4 (\pi n)^3 \frac{R^2 r_c}{L^3}$.

Then, in the leading approximation we obtain
\begin{IEEEeqnarray} {rCl} \label{eq:corrected_masses_odd}
m_2 = \left\{ 
\begin{aligned} 
&\frac{2 \pi n}{L} - 4 (\pi n)^3 \frac{R^2 r_c}{L^4}\,, \qquad \te{for } n \ll  n_{\te{max}}, &  \te{ (for } n \in \N)\,,&    \\
&\left. \begin{aligned} 
&\underbrace{\frac{\pi n}{L-R} + \frac{2}{L-R} \arctan f(n)}_{m_2^{\te{c}}}\,, \qquad \te{for }   n_{\te{min}}\ll n \ll n_*\,,   \\
&\frac{\pi n}{R}\,, 
\end{aligned} \right\} &   \te{ (for even integers  } n)\,,& 
 \end{aligned} \right.  \nonumber \\
\end{IEEEeqnarray} 
with
\begin{IEEEeqnarray}{rCl} \label{eq:corrected_masses_odd_approx}
m_2^{\te{c}} \simeq \left\{
\begin{aligned} 
&\tensor*[]{m}{^{\te{c}}_{2,a}} \equiv  \frac{\pi n}{L-R}  -\frac{\pi n}{L-R} \frac{R}{L}, & \qquad  n \ll n_\rho\,,&   \\
&\tensor*[]{m}{^{\te{c}}_{2,b}} \equiv \frac{\pi n}{L-R} + 2 \frac{L-R}{\pi n L r_c}\,, & \qquad  n \gg n_\rho\,.& 
 \end{aligned} \right. 
\end{IEEEeqnarray} 
Again, the form of the correction of $m_2^{\te{c}}$ is only valid up to $n \sim n_*$, with $n_*$ given in \eqref{eq:n_star}.

Note that the corrections, which we derived in \eqref{eq:corrected_masses_even} and \eqref{eq:corrected_masses_odd} improve the solutions for $m_\al$ as we approach the transition region from the IR transparent and the opaque regimes, respectively, but they are not valid at $m \sim r_c^{-1}$. Here, we cannot write down an asymptotic expansion for $m_\al$. However, if we extend the validity region of the solutions in \eqref{eq:corrected_masses_even} and \eqref{eq:corrected_masses_odd} all the way to $m \sim r_c^{-1}$, we will only introduce an error to the vacuum energy, which is of the order of $r_c^{-1}$. We can see this by looking at \eqref{eq:masses_rewritten}. For $m \sim r_c^{-1}$ the solutions will obviously be modified. However, the number of modes, which solve this equation, is still proportional to the number of times the left-hand side and the right-hand side crosses zero. In the limit $L > r_c \gg R$, this is of the order $\frac{L}{r_c}$, which is independent of $R$. Hence, the contribution to the energy from that region is 
\begin{equation}
\left( \sim \frac{1}{r_c} + \sim \frac{1}{L} \right) \frac{L}{r_c} = \mac{O}(1) \frac{L}{r_c^2} + \mac{O}(1) \frac{1}{r_c}\,, \label{eq:approx_explain}
\end{equation}
which deviates by $\mac{O}(1)$ factors from the true result. The first term cancels from the two sums $\tensor*[]{\Sigma}{^{\te{s}}_1}$ and $\tensor*[]{\Sigma}{^{\te{s}}_2}$, as can be seen explicitly in \eqref{eq:E_reg_sharp_2dim},\footnote{This is just an "artefact" from the fact that we split the sum there.} but the second term will contribute to the energy. Hence, correcting the masses around $r_c^{-1}$ would just change the $r_c^{-1}$-part of the vacuum energy by an $\mac{O}(1)$ numerical factor. Since this constant term will drop out in the expression of the Casimir force, we can safely use the expansions \eqref{eq:corrected_masses_even} and \eqref{eq:corrected_masses_odd} for $n \ll n_*$ ($m_\al \ll R^{-1}$). Although the corrections lose their validity as we approach $m_\al \sim R^{-1}$, they are suppressed as $\sim R/r_c$. Thus, in this section, we will consider the above expressions to be good approximate solutions (up to corrections $R/r_c$) of \eqref{eq:masses} for the full range of masses. However, we will see in the next section by performing a numerical analysis that the corrections for $n \gtrsim n_*$ will in fact also contribute, since they do not fall off fast enough at the lower limit. 

We further observe that the relevant correction to the Casimir energy (due to the "leakage") comes from the region $R < \lambda < r_c$. There, the correction of the even modes goes like $1/n$ and hence does not vanish fast enough, leading to a logarithmic contribution to the energy. The correction of the odd modes also has a contribution $1/n$ (for $n \gg n_\rho$), however, more importantly, it also has a contribution that peaks around $n \sim n_\rho$, where $m_2 \sim \rho^{-1}$. Hence, the energy will get a contribution $\rho^{-1}$, as we will explain below.

Let us now study the new contributions to \eqref{eq:E_reg_sharp_2dim} in detail. If we use the (corrected) even mass from \eqref{eq:corrected_masses_even} for the IR transparency region and plug it into $\tensor*[]{\Sigma}{^{\te{s}}_1}$, we find that \eqref{eq:sigma_sharp_1} acquires the additional contribution
\begin{equation*}
- \frac{1}{4 \pi r_c} + \ldots 
\end{equation*}
where the ellipses denote terms vanishing in the limit $a \to 0$, $L \to \infty$. However, as we explained before, the precise numerical factor of the term $\sim 1/r_c$ is only numerically calculable and not the one given above. The correction of the odd mass in the IR transparency region, \eqref{eq:corrected_masses_odd}, is much smaller than for the even modes. Since we did not expand the even masses up to that order, we have to use the leading approximation for the odd modes, if we want a consistent error estimate for $\tensor*[]{\Sigma}{^{\te{s}}_1}$.\footnote{If we would take into account this correction, $\Sigma_1$ would now contain a term $\propto \frac{R^2}{r_c^3}$.} Hence, there is no additional contribution to $\tensor*[]{\Sigma}{^{\te{s}}_1} $ from the odd modes. Thus, the energy from the IR transparency region is modified according to
\begin{equation*}
\tensor*[]{\Sigma}{^{\te{s}}_1} \to \Sigma_1 = \tensor*[]{\Sigma}{^{\te{s}}_1} + \mac{O}(1) \frac{1}{r_c} + \ldots\,.
\end{equation*}
We see that, as expected, the correction of the IR transparency modes due to the non-sharp transition at $m \sim r_c^{-1}$ only modifies the $r_c^{-1}$-term of the energy.

Next, we consider the correction of $\tensor*[]{\Sigma}{^{\te{s}}_2}$ coming from the even modes. We cannot just plug in the correction, which we found in \eqref{eq:corrected_masses_even}, and sum up to infinity, because the correction is only valid up to $n \sim n_*$. Hence, we have to divide the sum into two parts,\footnote{Note that the first sum now starts "2 steps" after the transition region, because the first step is taken care of by the odd modes. The same is true for the second sum. However, since individual modes contribute only with $\sim 1/L$ to the sum, in the continuous limit $L \to \infty$, these subtleties could just as well be ignored.} 
\begin{equation*}
\hal \sum \limits_ {\substack{n=n_{\te{min}} +2 \\ (\te{odd } n)}}^{n_*}  m_1^{\te{c}} \ee^{-m_1^{\te{c}} \frac{a}{\pi}} + \hal \sum \limits_ {\substack{n=n_* +2 \\ (\te{odd } n)}}^{\infty}  \frac{\pi n}{L-R} \ee^{-\frac{a n}{L-R}}\,.
\end{equation*}
Therefore, the correction to $\tensor*[]{\Sigma}{^{\te{s}}_2}$ will come entirely from the first part. Unfortunately, it cannot be summed exactly. However, we can find an asymptotic expansion around $a=0$ for this sum (we have used \texttt{Mathematica} for this). The new contribution is
\begin{equation*}
- \frac{1}{2 \pi r_c} \ln \frac{R}{2 r_c} + \dots \,.
\end{equation*}

For the  correction of $\tensor*[]{\Sigma}{^{\te{s}}_2}$ coming from the odd modes, we again have to note that the correction, which we found in \eqref{eq:corrected_masses_odd}, is only valid up to $n \sim n_*$. Furthermore, because it is the rather complicated function $\sim \arctan f$, an asymptotic expansion of the sum, involving this correction, is not available. Therefore, we will split the sum even further and use the two approximate expressions, given in \eqref{eq:corrected_masses_odd_approx}, in the respective regions. Hence, we have to evaluate the sums
\begin{equation*}
\underbrace{\hal \sum \limits_ {\substack{n= n_{\te{min}} +1 \\ (\te{even } n)}}^{n_\rho}  \tensor*[]{m}{^{\te{c}}_{2,a}} \ee^{-\tensor*[]{m}{^{\te{c}}_{2,a}} \frac{a}{\pi}}}_{I_a} + \underbrace{\hal \sum \limits_ {\substack{n= n_\rho+2 \\ (\te{even } n)}}^{n_*-1}  \tensor*[]{m}{^{\te{c}}_{2,b}} \ee^{-\tensor*[]{m}{^{\te{c}}_{2,b}} \frac{a}{\pi}}}_{I_b} + \hal \sum \limits_ {\substack{n= n_*+1 \\ (\te{even } n)}}^{\infty}  \frac{\pi n}{L-R} \ee^{-\frac{a n}{L-R}}\,,
\end{equation*}
where the new contributions to $\tensor*[]{\Sigma}{^{\te{s}}_2}$ will come solely from $I_a$ and $I_b$. From $I_a$ we find the new contributions:
\begin{IEEEeqnarray}{rCl} \IEEEyesnumber \IEEEyessubnumber* \label{eq:sigma_2_odd_correction_a}
I_a \supset \frac{L}{4 \pi \rho^2} + \frac{\sqrt{2}}{4 \rho} - \frac{2 + \pi}{4 \pi r_c} + \frac{R}{8 \pi r_c^2}\,. 
\end{IEEEeqnarray}
The sum $I_b$, as in the even case, cannot be evaluated exactly, so we again have to rely on an asymptotic expansion around $a=0$. We find the new contributions:
\begin{IEEEeqnarray}{rCl} \IEEEyessubnumber*  \label{eq:sigma_2_odd_correction_b}
I_b \supset - \frac{L}{4 \pi \rho^2}- \frac{\sqrt{2}}{4 \rho} + \frac{1}{4 \pi r_c} - \frac{1}{4 \pi r_c} \ln \frac{R}{2 r_c}\,.
\end{IEEEeqnarray}
Note that to derive expressions \eqref{eq:sigma_2_odd_correction_a} and \eqref{eq:sigma_2_odd_correction_b} we assumed that the two limiting values \eqref{eq:corrected_masses_odd_approx} are valid right up to $n \sim n_\rho$. Then, adding \eqref{eq:sigma_2_odd_correction_a} and \eqref{eq:sigma_2_odd_correction_b}, the terms proportional to $L/\rho^2$ and $1/\rho$ cancel, in the same way as the terms $\propto L/r_c^2$ canceled in \eqref{eq:E_reg_sharp_2dim} patching together $\Sigma_1^{\te{s}}$ and $\Sigma_2^{\te{s}}$. However, as explained in the paragraph around \eqref{eq:approx_explain} for the case of region $m \sim r_c^{-1}$, here also the masses at $m \sim \rho^{-1}$ should get $\mac{O}(1)$ corrections, since the exact form of $m_2^{\te{c}}$ here is $\frac{2}{L-R} \arctan f$. But now the number of modes contributing to that correction is $\sim L/\rho$, hence the expressions \eqref{eq:sigma_2_odd_correction_a} and \eqref{eq:sigma_2_odd_correction_b} will have terms $\propto \rho^{-1}$ that do not cancel. 

Thus, we find that the corrections of both the even and the odd masses modify $\tensor*[]{\Sigma}{^{\te{s}}_2}$ according to
\begin{equation*}
\tensor*[]{\Sigma}{^{\te{s}}_2} \to \Sigma_2 = \tensor*[]{\Sigma}{^{\te{s}}_2} + \mac{O}(1) \frac{1}{r_c} + \mac{O}(1) \frac{1}{\rho} - \frac{3}{4 \pi r_c} \ln \frac{R}{2 r_c} + \frac{R}{8 \pi r_c^2}   + \dots   .
\end{equation*}

The sum, taking care of the "inside" modes, $\tensor*[]{\Sigma}{^{\te{s}}_3}$ is unaffected by the correction of the modes in our leading approximation, hence $\tensor*[]{\Sigma}{^{\te{s}}_3}=\Sigma_3$. Finally, with the modification
\begin{equation*}
\tensor*[^{(2\te{D})}]{E}{^{\te{reg}}_{\te{s}}} \to \tensor*[^{(2\te{D})}]{E}{^{\te{reg}}} \equiv \Sigma_1 + \Sigma_2 +\Sigma_3 ,
\end{equation*}
the DGP analogue of \eqref{eq:casimir_energy_sharp_2dim}, in the first approximation, is
\begin{IEEEeqnarray}{rCl} 
\tensor*[^{(2\te{D})}]{E}{_{\te{C}}} &=&  \lim_{\substack{a \to 0 \\ L \to \infty}} \left( \tensor*[^{(2\te{D})}]{E}{^{\te{reg}}}   - \tensor*[^{(2\te{D})}]{E}{^{\te{reg}}_0}  \right) \nonumber \\
&=& - \frac{\pi}{24 R} +  \mac{O}(1) \frac{1}{r_c} + \mac{O}(1) \frac{1}{\sqrt{R r_c}}  - \frac{3}{4 \pi r_c} \ln \frac{R}{2 r_c} + \frac{3}{8 \pi} \frac{R}{r_c^2}  + \mac{O}\left(\frac{R^2}{r_c^3} \right) .  \label{eq:casimir_energy_2dim}
\end{IEEEeqnarray}

Let us make a couple of comments about this result. We see that the two terms, which were already present in the sharp transition approximation, $\propto 1/r_c$ and $\propto \frac{R}{r_c^2}$, are slightly modified by this more detailed resolution of the transition region, which is expected. However, note that the exact numerical prefactors are sensitive to our rough approximations. The constant term ($\propto 1/r_c$) depends on the precise (only numerically accessible) solutions at the transition region. The modification of the linear term comes entirely from $\tensor*[]{m}{^{\te{c}}_{2,a}}$, where the correction term is $\sim \frac{R}{r_c}$ smaller than the corresponding correction of the even modes, so it is possible that the numerical factor would change further, if we would take into account higher correction terms of the even modes.

More importantly, we found two new contributions to the Casimir energy, which are due to the fact that the masses in the region $r_c^{-1} \lesssim m \lesssim R^{-1}$ get corrections from the "DGP nature" of the branes. These corrections vanish only slowly in the limit $R \ll r_c$. In fact, we found that the correction of the masses of the odd modes even increases in the region $r_c^{-1} \lesssim m \lesssim \rho^{-1}$, before again falling off like $\propto n^{-1}$, where the characteristic scale of the turning point is $\rho=\sqrt{R r_c}$. The growing correction of the odd modes leads to a $\frac{1}{\sqrt{R r_c}}$-contribution, while the correction $\propto n^{-1}$ gives a logarithmic contribution.

This leads to the (novel) Casimir force
\begin{equation}
\tensor*[^{(2\te{D})}]{F}{_{\te{C}}} = - \frac{\pi}{24 R^2} + \mac{O}(1) \frac{1}{R \sqrt{R r_c}} \, + \tikzmark{A} \ \frac{3}{4 \pi} \frac{1}{R r_c} + \mac{O}\left(\frac{1}{r_c^2} \right)\,. \label{eq:casimir_force_2dim}
\begin{tikzpicture}[overlay, remember picture,shorten <=1mm,font=\footnotesize]
\draw[<-] ([yshift=0.5ex] pic cs:A) -- ++ (0,-0.6) node[below] {$\mac{O}(1)$ ?};
\end{tikzpicture}
\bigskip
\medskip
\end{equation}

We will see in the next section that the coefficient in front of the term $\propto \frac{1}{R r_c}$ gets further modified by an $\mac{O}(1)$ factor. Furthermore, we will find that the coefficient in front of the term $\propto \frac{1}{R \sqrt{R r_c}}$ is positive, thus leading to a weakening of the Casimir force. This result reflects the fact that the DGP branes are not just ideal boundary conditions (being completely transparent to soft modes and completely opaque to hard modes), but have a finite transition region, where they "try to keep" the hard modes, but still "leak" some of them. This affects the branes' ability to confine modes inside their interior and hence weakens the Casimir force as compared to the case of perfect conducting plates. Since this leakage effect turns out to be stronger than the effect we discovered in the previous section (leading to a constant increase of the Casimir force), the resulting Casimir force is weaker than in the case of ideal boundary conditions.

\subsection{Numerical analysis of the Casimir energy} \label{subsec:casimir_2dim_numerics} 

\begin{figure}[t]
\centering
    \includegraphics{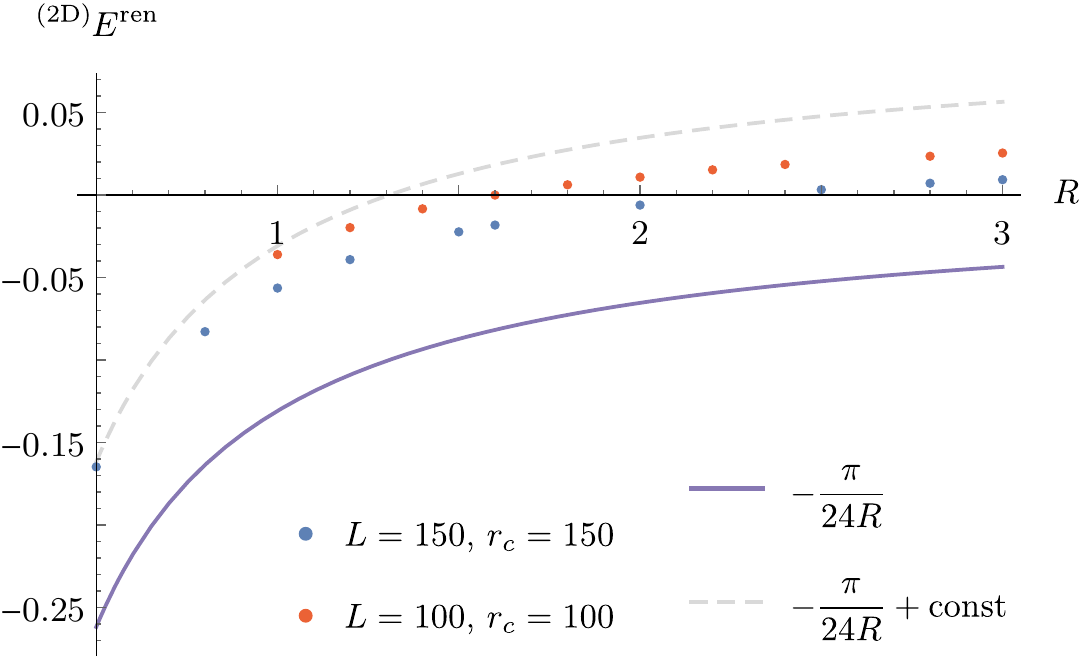}
    \caption[]{Casimir energy, calculated numerically (data points) for $a=0.1$ and two different sets of parameters $r_c$ and $L$. Compared to the standard result due to Dirichlet boundary conditions (purple solid curve) and the shifted version (gray dashed curve).}
    \label{fig:E_C_2dim}
   \end{figure}

The derivation of the Casimir force in the previous section relied on approximations of the quantized masses that have different forms in different regions. Since it is difficult to control the introduced errors in this way, in the present section we will  justify qualitatively the expression \eqref{eq:casimir_force_2dim} and obtain a positive sign for the leading correction by using numerical methods. Our starting point is to numerically solve \eqref{eq:masses} for several different values of $R$, $r_c$ and $L$. Since this equation has infinitely many solutions, we have to choose where we want to truncate them. We choose the largest mass in such a way that $m a  < 100$, because then the exponential factor $\exp (-m a/\pi)$ is sufficiently small such that the rest of the solutions does not contribute to the Casimir energy anymore. Then, using those numerical solutions, we calculate
\begin{equation}
\tensor*[^{(2\te{D})}]{E}{^{\te{ren}}} \equiv \tensor*[^{(2\te{D})}]{E}{^{\te{reg}}}   - \tensor*[^{(2\te{D})}]{E}{^{\te{reg}}_0} \label{eq:ren_energy_2dim}
\end{equation}
as a function of $R$. The result for $a=0.1$ and two different sets of parameters $r_c$ and $L$ is shown in Fig.~\ref{fig:E_C_2dim}.
In all of the following figures we express all quantities with dimension of length in some unit $l$, whereas the energy is expressed in units of $1/l$. We use dimensionless units by setting $l=1$. We see that the Casimir energy indeed approaches the form of the standard result, $- \frac{\pi}{24 R} $ (purple solid curve), but it also deviates from it. To see that the difference is not just a constant, but actually $R$-dependent, we also plotted the standard result with a constant shift (dashed gray curve).

\begin{figure}[t]
\centering
    \includegraphics{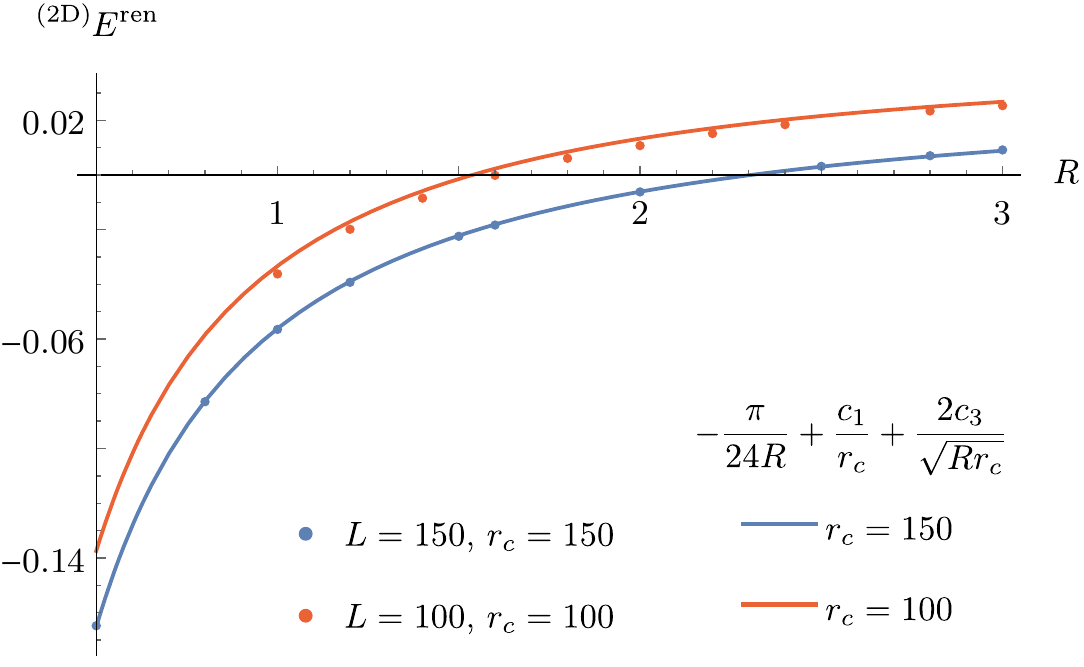}
    \caption[]{Casimir energy, calculated numerically (data points) for $a=0.1$ and two different sets of parameters $r_c$ and $L$. We fitted the analytical approximation $- \frac{\pi}{24 R} + \frac{c_1}{r_c} + \frac{2 c_3}{\sqrt{R r_c}}$, with free coefficients $c_1$ and $c_3$, to the numerical result for $r_c=150$ and $L=150$ (blue data points). The resulting function, with such determined coefficients $c_1 \simeq 3.32$ and $c_3 \simeq 0.32$, is plotted for $r_c=150$ (blue curve) and $r_c=100$ (red curve).}
    \label{fig:E_C_withfitleading_2dim}
   \end{figure}

Next, we compare the numerical result to our analytical approximation in Fig.~\ref{fig:E_C_withfitleading_2dim}.   
There, we fitted the leading three terms of \eqref{eq:casimir_energy_2dim} to the numerical result for $r_c=150$ and $L=150$ (blue data points), thereby obtaining the coefficients $c_1$ and $c_3$ numerically. Both are positive and $\mac{O}(1)$. We see that this function already reproduces very well the numerical result. The numerical result for $r_c=100$ and $L=100$ (red data points) is in slightly worse agreement, because for this value of $r_c$ the suppression $R/r_c$ is less pronounced and hence the higher order terms become more important. 

\begin{figure}[t]
\centering
    \includegraphics{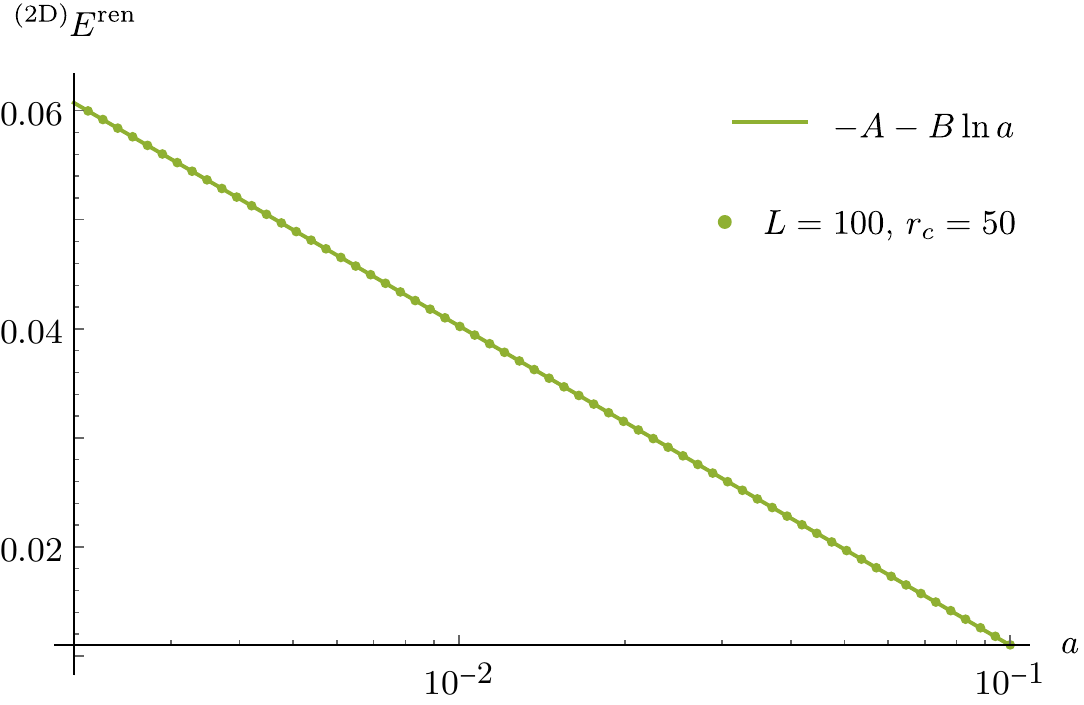}
    \caption[]{Log-linear plot of the numerical Casimir energy (green data points) for $R=1$, $L=100$ and $r_c=50$. The function $-A-B \ln a$, with free coefficients $A$ and $B$ is fitted to the data set. The resulting function, with such determined coefficients ($A,B > 0$), is plotted as green curve. We see that the Casimir energy depends logarithmically on $a$.}
    \label{fig:E_C_2dim_a}
   \end{figure}

So far, everything is as expected by the previous discussion. However, the numerical analysis shows one feature that was not visible in the analytical analysis. The Casimir energy seems to depend on the cut-off $a$ not just with powers $a,a^2,\ldots$ etc., which vanish in the limit $a \to 0$, but it has an ($\ln a$)-term. To see this, in Fig.~\ref{fig:E_C_2dim_a}
we plotted the Casimir energy (green points) for the fixed value $R=1$ as a function of $a$. Comparing it to the function $-A - B \ln a$ (green curve), we see that the logarithmic dependence is very robust (for this range of $a$, the terms with positive powers of $a$ are subleading). We suspect that a term proportional to $\frac{1}{r_c} \ln \frac{a}{R}$ arises in \eqref{eq:casimir_energy_2dim} because of the following reason. In our calculation of $\Sigma_2$ we approximated the correction to the even masses in the region $r_c^{-1} \lesssim m \lesssim R^{-1}$ as $\propto 1/n$. Therefore, the contribution to $\Sigma_2$ from that correction is 
\begin{equation*}
 \sim \frac{1}{r_c} \sum_{n \sim \frac{L}{r_c}}^{n \sim \frac{L}{R}} \frac{1}{n} \ee^{-\frac{a n}{L-R}} \hspace{-10pt} \underset{\te{\shortstack{$\bigg\uparrow$ \\ in the limit \\ $L \to \infty$}}}{\sim} \hspace{-10pt} \frac{1}{r_c} \left( \ln \frac{a}{R} - \ln \frac{a}{r_c} \right) = - \frac{1}{r_c} \ln \frac{R}{r_c}\,,
\end{equation*}
so the $a$-dependence drops out. Then, for the region $m \gtrsim R^{-1}$, we neglected the correction to the mass and hence did not get any further contribution. However, if we would derive a correction term there as well, it is expected that we would generate a term $\propto \frac{1}{r_c} \ln \frac{a}{R}$, coming from the lower limit of $m \sim R^{-1}$. A similar argument holds for the odd masses.

\begin{figure}[t]
\centering
    \includegraphics{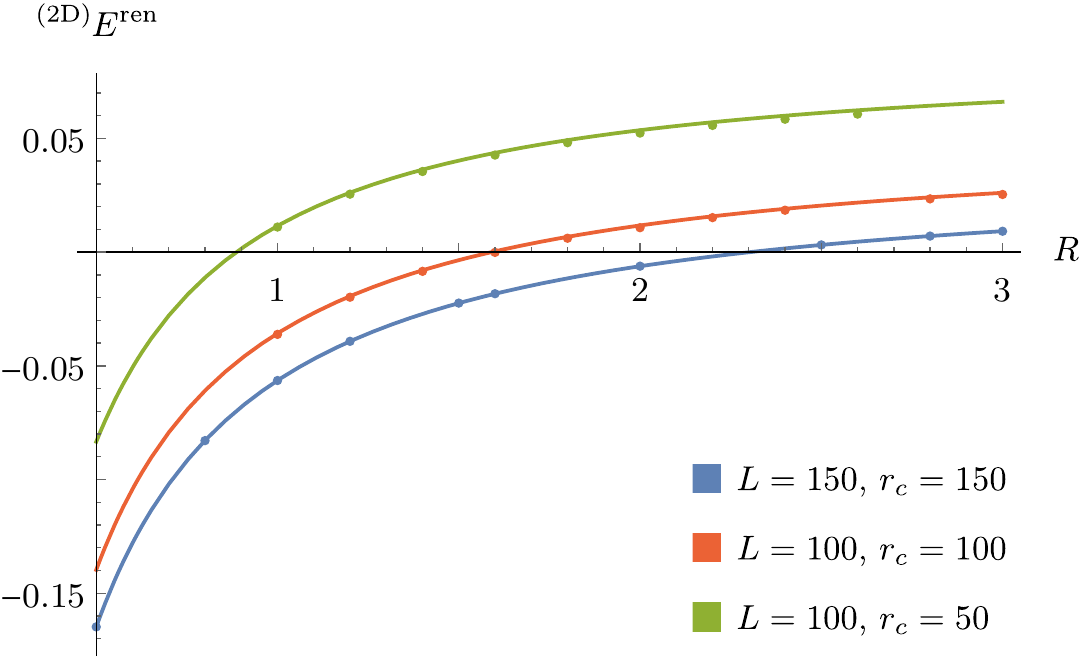}
    \caption[]{Casimir energy, calculated numerically (data points) for $a=0.1$ and three different sets of parameters $r_c$ and $L$. We fitted the analytical approximation $- \frac{\pi}{24 R} + \frac{c_1}{r_c} + \frac{c_2}{L} + \frac{2 c_3}{\sqrt{R r_c}}- \frac{c_4}{r_c} \ln \frac{a}{R}$, with free coefficients $c_i$, to the numerical result for $r_c=150$ and $L=150$ (blue data points). The resulting function, with such determined coefficients $c_1 \simeq 0.50$, $c_2 \simeq 0.48$, $c_3 \simeq 0.37$ and $c_4 \simeq 0.51$, is plotted for $r_c=150$ (blue curve), $r_c=100$ (red curve) and $r_c=50$ (green curve).}
    \label{fig:E_C_withfitlog_2dim}
   \end{figure}

In fact, our numerical analysis shows that the term $\propto \frac{1}{r_c} \ln \frac{a}{R}$ enters with a much larger coefficient than the term $\propto \frac{1}{r_c} \ln \frac{2 r_c}{R}$, which we found in \eqref{eq:casimir_energy_2dim}. In Fig.~\ref{fig:E_C_withfitlog_2dim}
we show how the $\frac{1}{r_c} \ln \frac{a}{R}$-term improves the numerical fit to the data points. Even, if we include another data set (green points) with an even less pronounced suppression $R/r_c$, the fit is very good.\footnote{Note that now we had to include a term $\propto 1/L$ in the plot of Fig.~\ref{fig:E_C_withfitlog_2dim}. In the plot of Fig.~\ref{fig:E_C_withfitleading_2dim} we absorbed it into the term $\propto \frac{1}{r_c}$, since we had $r_c=L$. However, for the data points with $r_c \neq L$ the curve would get a constant shift, if we would not take this into account. Of course, other than that, this term is not relevant, since it vanishes in the limit $L \to \infty$.}

\begin{figure}[t]
\centering
    \includegraphics[width=0.7\linewidth]{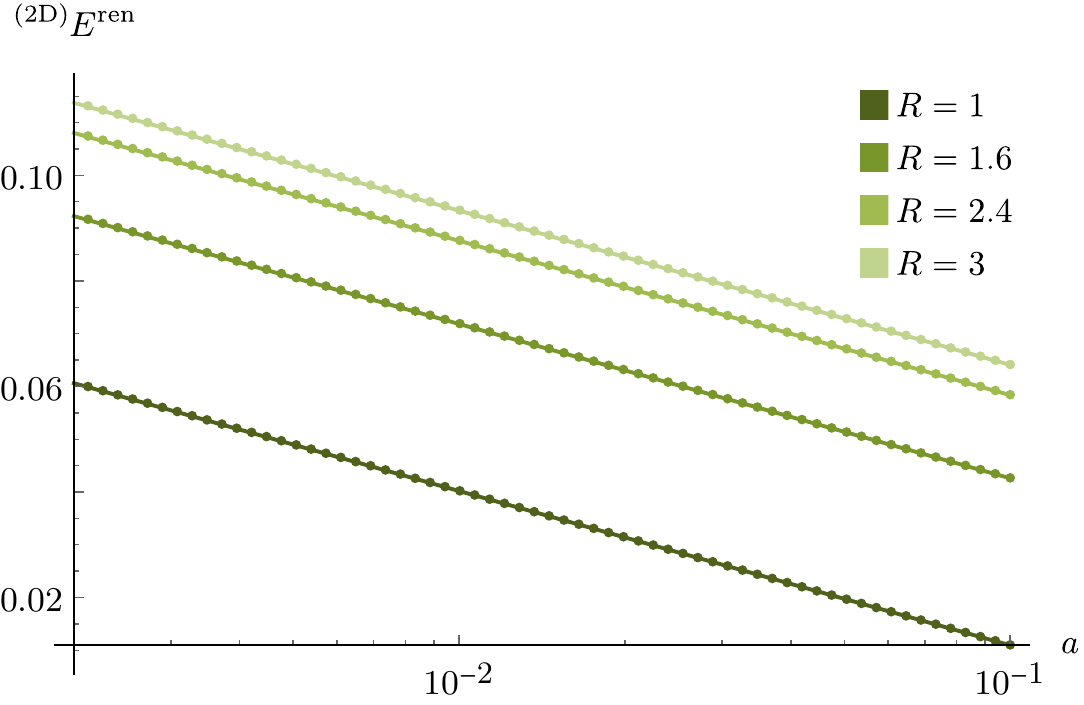}
    \caption[]{This log-linear plot shows the functions $-A-B \ln a$ (solid curves) that are fitted to the numerical results for $L=100$ and $r_c=50$ (data points) for different values of $R$. The slopes $B \simeq 0.0127$ are equal for all curves.}
    \label{fig:E_C_2dim_a_indep}
   \end{figure}

Now, at the first glance the logarithmic term seems problematic, because it depends on $a$ and does not vanish in the limit $a \to 0$. However, the force is the actual observable and for that the cut-off parameter drops out. Indeed, the term $\propto \frac{1}{r_c} \ln \frac{a}{R}$ will contribute to the force in the same way as the 3rd term in \eqref{eq:casimir_force_2dim}, modifying its numerical factor. To prove that the term $\propto \frac{1}{r_c} \ln \frac{a}{R}$, which we found numerically, does not have an $R$-dependent coefficent (which would invalidate the previous argument), we show in Fig.~\ref{fig:E_C_2dim_a_indep}
the results for different values of $R$. The fitted curves (solid lines) all have the same slope $B$. Only the off-sets, which of course depend on $R$, differ.

After we established the correctness of \eqref{eq:casimir_force_2dim} numerically and found the magnitudes and signs of the coefficients, we can finally state the Casimir force as
\begin{equation}
\tensor*[^{(2\te{D})}]{F}{_{\te{C}}} = - \frac{\pi}{ 24 R^2} + \frac{c_3}{R \sqrt{R r_c}} - \frac{c_4}{R r_c} + \mac{O}\left(\frac{1}{r_c\sqrt{R r_c}} \right)\,,
\end{equation}
with the coefficients $c_3 \simeq 0.37$ and $c_4 \simeq 0.51$.

Thus, we find indeed that, in the presence of the IR transparency region, the Casimir force becomes weaker due to the "leakage" of hard modes.

\section{Casimir effect in 4+1 dimensions} \label{sec:casimir_5dim}

After we have analyzed the properties of the given system in 2D, which led to the novel Casimir force, we can now turn our attention to the system in (the full) 5 dimensions. In principle, we do not expect a qualitative difference from our previous result, since the branes that produce the effective boundary conditions are still codimension-one objects, and the discrete masses of the KK modes are still given by \eqref{eq:masses}. However, the sum that has to be evaluated in order to get the Casimir energy is much more involved in 5D (see \eqref{eq:E_reg_5dim}) than in 2D (see \eqref{eq:E_reg_2dim}). For this reason, we are not able to completely reproduce the analytic analysis and the quantitative results of the previous sections, but have to rely on numerical methods. We will explain these matters in the following.

Let us start with expression \eqref{eq:E_reg_5dim} and note that we can perform the integration, which leads to
\begin{equation} 
E^{\te{reg}} = -\frac{1}{4} \frac{\del}{\del a} \sum \limits_{\alpha,m}  \frac{m_{\al}^2}{a} K_2\left(\frac{a m_{\al}}{\pi} \right),  \label{eq:E_reg_5dim_integrated}
\end{equation}
where $K_2$ is the modified Bessel function of the 2nd kind.

Now, we would like to calculate an analogue of \eqref{eq:E_reg_sharp_2dim} in the (toy model) approximation of a sharp transition, however the sum \eqref{eq:E_reg_5dim_integrated}, with the masses given by \eqref{eq:infra_trans_masses} and \eqref{eq:opaque_masses}, cannot be solved analytically. Moreover, we were not able to find an asymptotic expansion of this sum.

However, we will show in the next section that in the extreme case $r_c \to \infty$, where $\tensor*[]{\Sigma}{^{\te{s}}_1}$ vanishes (because there is no IR transparency regime anymore), we can use \eqref{eq:E_reg_5dim_integrated} to derive the standard Casimir energy, proving that also in 5D the DGP branes, which provide only effective boundary conditions in the finite $r_c$ case, approach Dirichlet boundary conditions in the limit of infinite $r_c$.

\subsection{Analytical result for the Casimir force in the limit $r_c \to \infty$} 

First, note that the Hamiltonian in the 5D case does include the zero-mode ($m_1=0$), because even if it is constant along the extra dimension, the oscillations along the transverse directions still contain energy ($\omega_{m=0,\al=1}(\vec{k}) \neq 0$). Also, $m_1=0$ is still a solution of \eqref{eq:masses}, even in the limit $r_c \to \infty$. So, separating out this zero-mode and using the masses \eqref{eq:opaque_masses}, valid in the limit $r_c \to \infty$, we get
\begin{equation} 
E^{\te{reg}}(r_c \to \infty) = \frac{3}{2} \frac{\pi^2}{a^4} -\frac{\pi^2}{4} \frac{\del}{\del a} \left[ \frac{1}{a} \left( \frac{1}{d_1^2} S(d_1) + \frac{1}{d_2^2} S(d_2) \right) \right]\,,   \label{eq:E_reg_5dim_large_rc}
\end{equation}
with
\begin{equation} 
S(d_i) \equiv \sum_{n=1}^{\infty} n^2 K_2\left(\frac{a n}{d_i} \right)\,, \qquad d_1 = L-R\,, \qquad d_2=R\,. \label{eq:bessel_sum}
\end{equation}

The evaluation of $S(d_i)$ is performed in appendix \ref{app:sum}, where we find
\begin{equation*} 
E^{\te{reg}}(r_c \to \infty) = \frac{3 \pi^3}{2} \frac{L}{a^5} - \frac{\pi^2 \zeta'(-4)}{32} \frac{1}{R^4} - \frac{\pi^2 \zeta'(-4)}{32} \frac{1}{(L-R)^4}   +\mac{O} \left( \frac{a^2}{d_i^2} \right)\,,   
\end{equation*}
with $\zeta'$ the derivative of the Riemann zeta function. In order to renormalize this expression, we have to once again subtract the vacuum energy for a system without branes. The latter is conveniently derivable from \eqref{eq:E_reg_5dim_integrated} for masses \eqref{eq:plane_masses}. Thus, we get
\begin{equation*} 
E^{\te{reg}}_0  = E^{\te{reg}}(r_c \to 0) = \frac{3}{2} \frac{\pi^2}{a^4} -\frac{\pi^2}{2} \frac{\del}{\del a} \left( \frac{1}{a}  \frac{1}{d_3^2} S(d_3)  \right),   
\end{equation*}
with $S(d_3)$ again given by \eqref{eq:bessel_sum}, for $d_3=L/2$. Then, again using \eqref{eq:res_sum_result}, we find
\begin{equation*} 
E^{\te{reg}}_0 = \frac{3 \pi^3}{2} \frac{L}{a^5} - \pi^2 \zeta'(-4) \frac{1}{L^4}  +\mac{O} \left( \frac{a^2}{L^2} \right) .   
\end{equation*}
With this, we obtain the Casimir energy (per unit 3-volume) as
\begin{IEEEeqnarray}{rCl} 
E_{\te{C}}(r_c \to \infty) &=& \lim_{\substack{a \to 0 \\ L \to \infty}} \left( E^{\te{reg}}(r_c \to \infty)   - E^{\te{reg}}(r_c \to 0) \right) \nonumber \\
&=& - \frac{\pi^2 \zeta'(-4)}{32} \frac{1}{R^4}  \label{eq:casimir_energy_5dim_limit}
\end{IEEEeqnarray}
and the Casimir force (per unit 3-volume) as\footnote{We used here $\zeta'(-4)=\frac{3 \zeta(5)}{4 \pi^4}$.}
\begin{equation} 
F_{\te{C}}(r_c \to \infty) = - \frac{3 \zeta(5)}{32 \pi^2} \frac{1}{R^5} .
\end{equation}
Notice that this is the same result as one would obtain for a system with a real, massless, 5D scalar field and two parallel "conducting" plates (with Dirichlet boundary conditions) separated along the 5th dimension \cite{Ambjorn1983}. However, we calculated this result in the DGP setup using the limit $r_c \to \infty$ (after an explicit dimensional reduction), proving that the DGP branes can indeed lead to effective Dirichlet boundary conditions, which give rise to the standard Casimir force in the limit of a vanishing IR transparency region.

\subsection{Numerical derivation of the Casimir force}

Since we are not able to derive the 5D analogues of \eqref{eq:casimir_force_sharp_2dim} and \eqref{eq:casimir_force_2dim}, in the 5D system we have to rely entirely on numerical methods to show that the Casimir force gets weakened due to the leakage of the modes. 

In order to calculate \eqref{eq:E_reg_5dim_integrated} numerically, we rewrite it as
\begin{equation} 
E^{\te{reg}} =\frac{3 \pi^2}{2 a^4} + \frac{1}{8 \pi} \frac{1}{a^2} \sum \limits_{\alpha,m>0} m_\al^2 \left[ 2 \pi K_2\left(\frac{a m_{\al}}{\pi} \right) + a m_\al \left(K_1\left(\frac{a m_{\al}}{\pi} \right) + K_3\left(\frac{a m_{\al}}{\pi} \right) \right) \right] . \label{eq:E_reg_5dim_integrated_rewritten}
\end{equation}
The vacuum energy without branes, which we have to subtract from \eqref{eq:E_reg_5dim_integrated_rewritten}, is obtained, if we insert the masses \eqref{eq:plane_masses} in the above equation (where the zero-mode has already been separated out), leading to
\begin{equation*} 
E^{\te{reg}}_0 =\frac{3 \pi^2}{2 a^4} + \frac{\pi}{L^2 a^2} \sum_{n=1}^{\infty} n^2 \left[ 2 \pi K_2\left(\frac{2 a n}{L} \right) + a m_\al \left(K_1\left(\frac{2 a n}{L} \right) + K_3\left(\frac{2 a n}{L} \right) \right) \right] .  
\end{equation*}

\begin{figure}[t]
\centering
    \includegraphics[width=0.7\linewidth]{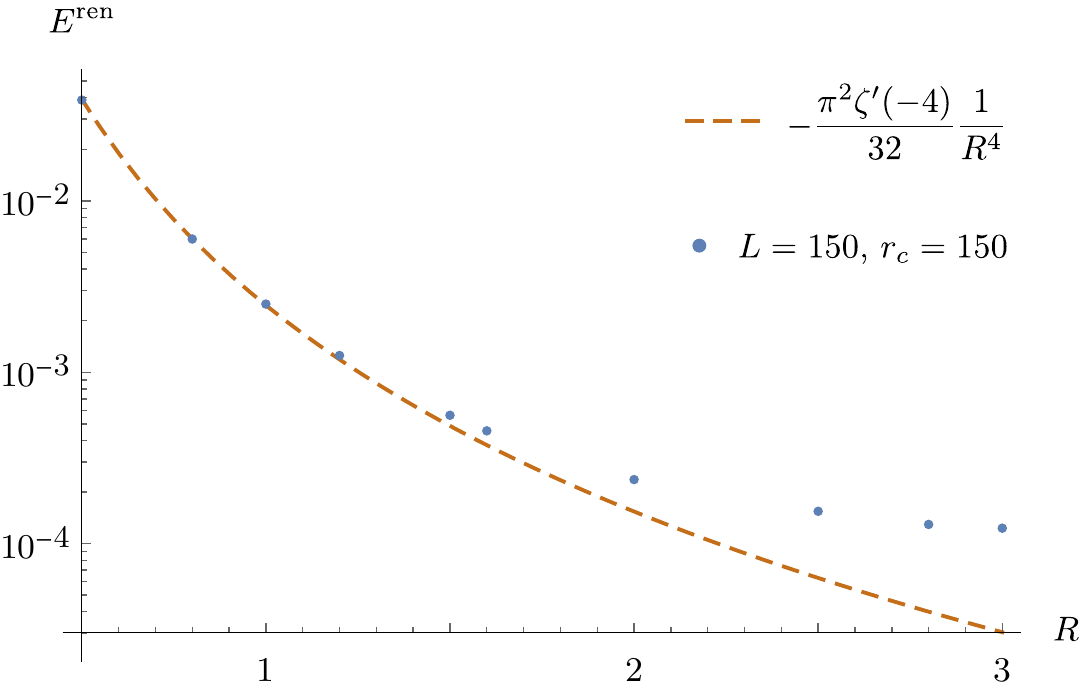}
    \caption[]{Log-plot of the Casimir energy, calculated numerically (blue data points) for $a=0.1$, $r_c=150$ and $L=150$. The brown dashed curve represents the standard (5D) Casimir energy (due to Dirichlet boundary conditions). We plotted the absolute values.}
    \label{fig:E_C_5dim}
   \end{figure}

Fig.~\ref{fig:E_C_5dim} 
shows the result for calculating   
\begin{equation}
E^{\te{ren}} \equiv E^{\te{reg}} - E^{\te{reg}}_0 \label{eq:ren_energy_5dim}
\end{equation}
numerically (for $r_c=150$, $L=150$ and $a=0.1$), where we again shifted the result by an $a$-dependent constant (i.e.\ $R$-independent). We see that the standard Casimir energy (i.e.\ \eqref{eq:casimir_energy_5dim_limit}) changes faster than our derived one. Hence, analogous to the 2D case, we find that the DGP branes weaken the resulting Casimir force.

\begin{figure}[t]
\centering
    \includegraphics[width=0.7\linewidth]{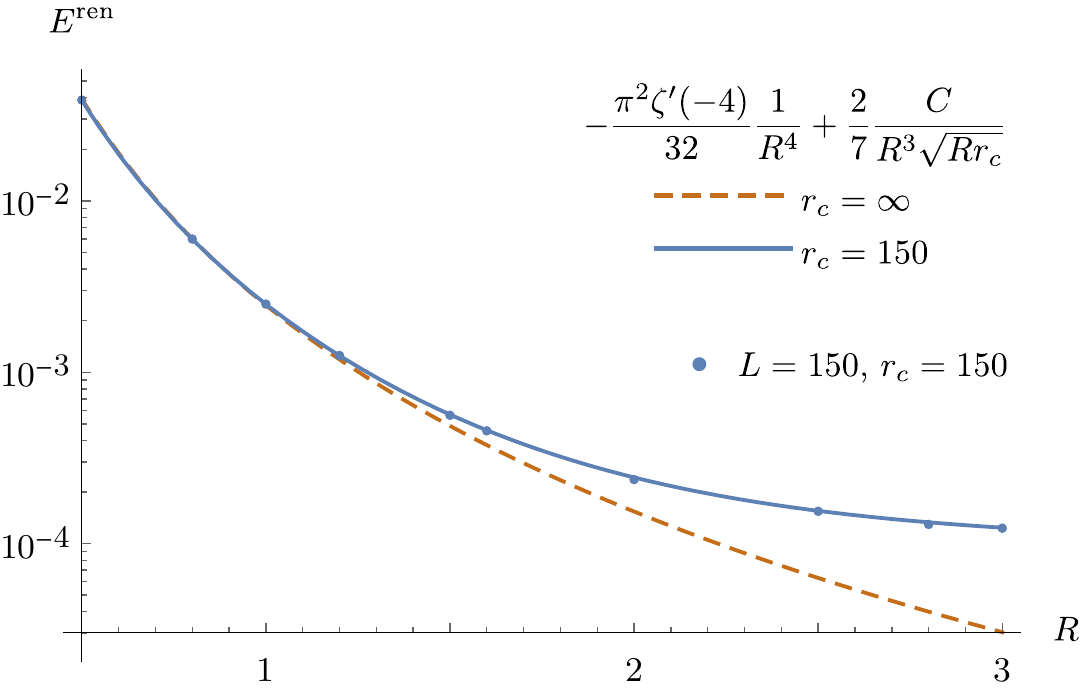}
    \caption[]{This log-plot shows the Casimir energy (blue data points) for the parameters $a=0.1$, $r_c=150$ and $L=150$. Also shown is the plot of the function, stated in the plot legend, that was fitted to the data points thereby determining the coefficient $C$. We suppress here the constant shift, which we had to apply to the curve in order to fit the points. The brown, dashed curve represents the standard (5D) Casimir energy (due to Dirichlet boundary conditions). We plotted the absolute values.}
    \label{fig:E_C_5dim_withfitleading}
   \end{figure}

If we now guess the first (non-constant) correction by analogy with the 2D case, \eqref{eq:casimir_energy_2dim}, we would suspect a term $\sim \frac{1}{R^3 \rho}$. We can convince ourselves by looking at Fig.~\ref{fig:E_C_5dim_withfitleading}
that this correction gives indeed a good fit to the data. Hence, we finally establish the 5D Casimir force as
\begin{equation}
\tensor*[]{F}{_{\te{C}}} = - \frac{3 \zeta(5)}{32 \pi^2} \frac{1}{R^5}  + \frac{C}{R^4 \sqrt{R r_c}} + \mac{O}\left(\frac{1}{R^4 r_c} \right) ,
\end{equation}
with the positive coefficient $C \simeq 2.6 \cdot 10^{-3}$. The smallness of this coefficient should not surprise us, since the coefficient of the leading term is already as small as $\frac{3 \zeta(5)}{32 \pi^2} \simeq 9.8 \cdot 10^{-3}$.\footnote{The coefficients in front of the (ordinary) Casimir force usually decrease rapidly with increasing dimension \cite{Ambjorn1983}.}

\section{Conclusion and summary} \label{sec:conclusion}

We have investigated and established the existence of the Casimir effect in a system without fixed (Dirichlet) boundary conditions, but with surfaces that have the property to suppress the high energy modes of a quantum field while being transparent to the low energy modes. As a particular example of such a system, we have studied the DGP model with two parallel 3-branes in a 5D spacetime, where the DGP branes provide "effective" boundary conditions for the 5D quantum field.

Furthermore, we have shown that the just described phenomenon, called IR transparency, has a profound implication for the arising Casimir force: it is weakened. 

In order to analyze quantitatively how the IR transparency affects the Casimir force, we have studied in detail the 2D version of the system (two parallel 0-branes separated along the extra dimension) and found that the IR transparency phenomenon has in fact two opposite contributions to the Casimir force. First, in the case of sharply separated regimes of "opaque" and "IR transparent" branes, the Casimir force increases due to the exclusion of IR modes from the spectrum of vacuum fluctuations between the branes. Second, since the DGP branes distinguish between hard and soft modes in a smooth manner, some of the hard modes "leak" out of the interior of the branes, thereby weakening the Casimir force. It turns out that the latter effect dominates and hence the resulting Casimir force decreases.

We have analytically derived the qualitative expressions of the leading correction terms to the Casimir force in the 2D case and justified the expressions by numerical analysis for both the 2D and 5D system. Regardless of the dimensionality we find that the corrections are suppressed by increasing powers of $\rho$, where $\rho \equiv \sqrt{R r_c}$ is a new distance scale (geometric mean between the separation distance of the branes and the DGP cross-over scale) arising in this system. The similarity between systems of different dimensions is not surprising, since in both cases the effect comes from codimension-one DGP branes separated along the extra dimension. It is perhaps interesting to note that the same scale $\rho$ has been found in Ref.~\cite{Warkentin:2019caf} in a very different context. There, it marked the transition between the standard DGP like and a novel, weaker behavior of the gravitational force between two sources (located on the opposite branes).

We have performed our analysis for a massless real scalar field with localized kinetic terms on parallel surfaces, thereby showing that the derived effect is very generic. It is beyond the scope of this paper to repeat the analysis for spin-1 and spin-2 fields. However, as we explained in section \ref{sec:quantizing}, we do not expect the results to change except for numerical $\mac{O}(1)$ factors due to the increased number of degrees of freedom. This should be verified in a future work.

Since the emergent Casimir force --- along with its corrections --- is a signature for braneworld scenarios with parallel DGP branes, it might be interesting for future research to further investigate how the presence of this force affects those scenarios. Moreover, since this effect should also be present for a large number of setups, including our 3+1 dimensional world, it is worthwhile to investigate, if there are materials that allow for surfaces with a sufficiently low scale $r_c$ such that an observation of this modified Casimir force is possible.

Finally, the presence of the derived effect shows that it is natural to construct "effective" boundary conditions for the graviton in the described way, thereby providing a mechanism to produce the gravitational Casimir effect and the means to probe the quantization of the gravitational field, which otherwise eludes an experimental access. However, we want to point out that the parameter $r_c$ for a quasi-localization of gravitons on surfaces in our 4D world should be minuscule. But then, the above result is not valid anymore, since we worked in the limit $r_c \gg R$ and implicitly assumed that the width of the branes is much smaller than $R$ (to be able to use the idealization of zero width). Therefore, to trust our result, $r_c$ has to be much larger than the width of the surfaces --- a situation which is probably not realized for realistic materials in two-dimensional surfaces in our world.
\clearpage

\appendix

\section{Mode functions} \label{app:mode}

In order to find the solutions $\psi_{m,\alpha}(y)$ of \eqref{eq:schroedinger}, we have to solve the harmonic oscillator equation in the three regions I, II and III (see Figure~\ref{fig:setup}) and then match the solutions at the boundaries of those regions (this is the standard procedure for Schroedinger type equations with delta-function potentials). In particular, the matching conditions (at the branes) between the regions I and II and the regions II and III are: continuity of $\psi_m(y)$ at $y=-\frac{R}{2}$ and $y=\frac{R}{2}$, discontinuity of $\frac{\dd \psi_m(y)}{\dd y}$ at $y=-\frac{R}{2}$ and $y=\frac{R}{2}$.\footnote{$\lim \limits_{\epsilon \to 0} \frac{\dd \psi_m(y)}{\dd y}\Bigr|^{\pm \frac{R}{2} +\epsilon}_{\pm \frac{R}{2}-\epsilon}+m^2 r_c \psi_m(\pm \frac{R}{2})=0$.} This calculation leads to
\begin{IEEEeqnarray*}{rCl} 
\psi_m(y) = \left\{ 
\begin{aligned} 
&A_m \ee^{\iu m y} + B_m \ee^{- \iu m y} , & - \frac{L}{2} < y<-\frac{R}{2}& , \\ 
&C_m \ee^{\iu m y} + D_m \ee^{- \iu m y} , & - \frac{R}{2} < y< \frac{R}{2}& , \\
&E_m \ee^{\iu m y} + F_m \ee^{- \iu m y} , &  \frac{R}{2} < y< \frac{L}{2}& ,
\end{aligned} \right. 
\end{IEEEeqnarray*}
where 4 of the 6 integration constants are fixed as
\begin{IEEEeqnarray*}{rCl} 
A_m &=& C_m \left(1-  \frac{\iu m r_c}{2} \right) -  D_m \frac{\iu m r_c}{2} \ee^{\iu m R}, \\ 
B_m &=& C_m \frac{\iu m r_c}{2} \ee^{- \iu m R} + D_m \left(1+ \frac{\iu m r_c}{2}\right) , \\ 
E_m &=& C_m \left(1 + \frac{\iu m r_c}{2} \right) +  D_m \frac{\iu m r_c}{2} \ee^{- \iu m R},\\
F_m &=& - C_m \frac{\iu m r_c}{2} \ee^{\iu m R} + D_m \left(1- \frac{\iu m r_c}{2}\right) ,
\end{IEEEeqnarray*}
with two arbitrary integration constants $C_m$ and $D_m$.

The conditions of periodicity, $\psi_m(\frac{L}{2})=\psi_m(-\frac{L}{2})$ and $\del_y \psi_m|_{\frac{L}{2}}=\del_y \psi_m|_{-\frac{L}{2}}$, lead to the quantization of the masses according to
\begin{equation}
\ee^{\iu m L} = \frac{b^*_\al(m)}{b_\al(m)}\,, 
\qquad \te{for} \ \left\{
\begin{aligned} 
&C_m = D_m , &  \al=1& , \\ 
&C_m = - D_m , & \al=2& ,
\end{aligned} \right. \label{eq:app_masses}
\end{equation}
with
\begin{equation}
b_{1,2}(m) = \hal \left( 1+ \frac{\iu m r_c}{2} \left(1 \pm \ee^{- \iu m R} \right) \right) . \label{eq:b}
\end{equation}
Notice that the solutions split into even ($\al=1$) and odd ($\al=2$). The mass quantization equation \eqref{eq:app_masses} can be rewritten as
\begin{equation}
\tan \frac{m L}{2} =- \frac{\Im(b_\al(m))}{\Re(b_\al(m))}\,, \label{eq:masses_app}
\end{equation}
which leads to the expression given in \eqref{eq:masses}.

Putting everything together, the (real-valued) solutions for the mode functions are
\begin{IEEEeqnarray}{rCl} \label{eq:modes_fct_even}
\psi_{m,1}(y) &=& \left\{ 
\begin{aligned} 
&C_1(m_1) b_1(m_1) \ee^{- \iu m_1 y} +c.c. , & \quad  -\frac{L}{2} < y<-\frac{R}{2}& , \\
&C_1(m_1) \cos{m_1 y} , & \quad  -\frac{R}{2} < y<\frac{R}{2}& , \\  
&C_1(m_1) b_1(m_1) \ee^{ \iu m_1 y} +c.c. , & \quad  \frac{R}{2} < y< \frac{L}{2}& ,
\end{aligned} \right. \qquad \te{(even)}
\end{IEEEeqnarray} 
and
\begin{IEEEeqnarray}{rCl} \label{eq:modes_fct_odd}
\psi_{m,2}(y) &=& \left\{ 
\begin{aligned} 
&\iu C_2(m_2) b_2(m_2) \ee^{- \iu m_2 y} +c.c. , & \quad  -\frac{L}{2} < y<-\frac{R}{2}& , \\
&C_2(m_2) \sin{m_2 y} , & \quad  -\frac{R}{2} < y<\frac{R}{2}& , \\  
&-\iu C_2(m_2) b_2(m_2) \ee^{ \iu m_2 y} +c.c. , & \quad  \frac{R}{2} < y< \frac{L}{2}& ,
\end{aligned} \right. \qquad \te{(odd)}
\end{IEEEeqnarray} 
with $b_\al(m)$ and $m_\al$ given by \eqref{eq:b} and the solutions of \eqref{eq:masses_app}, respectively.

The integration constants $C_{\al}(m_\al)$ can be fixed by requiring a normalization according to \eqref{eq:completeness}. We find\footnote{As explained in section~\ref{sec:quantizing}, there is no zero-mode for $\al=2$.}
\begin{equation*}
C_{\al}(m_\al)= \frac{1}{\sqrt{2}} {\left( |b_\al(m_\al)|^2 (L-R)+\frac{R}{4}+\frac{\Im(b_\al(m_\al))}{m_\al} \right)}^{-1/2}\,, \qquad \te{for } m_1 \neq 0\,,
\end{equation*}
and
\begin{equation*}
C_1(0) = \frac{1}{\sqrt{2 r_c + L}}\,, \qquad \te{for } m_1 = 0\,.
\end{equation*}

\section{Evaluation of the Bessel function sum} \label{app:sum}

In order to evaluate the sum \eqref{eq:bessel_sum}, we will closely follow Ref.~\cite{paris2018evaluation}, where several asymptotic expansions of Bessel function sums similar to ours are derived. We start by defining the new variable $\tau := \frac{a}{d_i}$ and taking the Mellin transform of
\begin{equation*}
S\left( \frac{a}{d_i} \right) = S(\tau) = \sum_{n=1}^{\infty} n^2 K_2\left(\tau n\right) ,
\end{equation*}
which leads to
\begin{IEEEeqnarray}{rCl} \label{eq:bessel_int}
\widetilde{S}(s) &=& \int_0^{\infty} \dd \tau \, \tau^{2 s-1} S(\tau) \nonumber \\
&=& \sum_{n=1}^{\infty} n^2 \underbrace{\int_0^{\infty} \dd \tau \, \tau^{2 s-1}  K_2 \left(\tau n\right)}_{J} 
\end{IEEEeqnarray} 
that converges for $\Re(s) > s_0$, with $s_0$ specified below. The integral, $J$, in \eqref{eq:bessel_int} can be easily performed, giving
\begin{equation*}
J = \frac{1}{4} 2^{2 s} \Gamma(s-1) \Gamma(s+1) n^{-2 s} ,
\end{equation*}
with the Gamma function $\Gamma$. Now we can take the sum 
\begin{equation*}
\sum_{n=1}^{\infty} n^{2-2 s} = \zeta(2 s-2) ,
\end{equation*}
where $\zeta$ is the Riemann zeta function, and find
\begin{equation*}
\widetilde{S}(s)= \frac{1}{4} 2^{2 s} \Gamma(s-1) \Gamma(s+1) \zeta(2 s-2) .
\end{equation*}
Finally, we take the inverse Mellin transform
\begin{IEEEeqnarray}{rCl} \label{eq:inverse_trafe}
S(\tau) &=& \frac{1}{\pi \iu} \int \limits_{s_0 - \iu \infty}^{s_0 + \iu \infty} \dd s \, \tau^{-2 s} \widetilde{S}(s) \nonumber \\
&=& \frac{1}{4 \pi \iu} \int \limits_{s_0 - \iu \infty}^{s_0 + \iu \infty} \dd s \, \tau^{-2 s} 2^{2 s} \Gamma(s-1) \Gamma(s+1) \zeta(2 s-2) .
\end{IEEEeqnarray} 
The integrand in \eqref{eq:inverse_trafe} has an infinite set of poles, coming from both the Riemann zeta function (pole at $s=\frac{3}{2}$) and the Gamma functions. The poles at $s=\frac{3}{2},1,0$ are simple, while the ones at $s=-1,-2,\ldots$ are double poles. We now see that $s_0=\frac{3}{2}$, so the integration contour in \eqref{eq:inverse_trafe} has to be on the right-hand side of this in order for $S(\tau)$ to converge. Using the residue theorem, we can rewrite the above expression as
\begin{equation}
S(\tau) = \hal \sum_{s_i} \te{Res}(f(s),s_i) , \label{eq:res_sum}
\end{equation}
where we sum over all the residues of
\begin{equation*}
f(s) = 4^s \tau^{-2 s} \Gamma(s-1) \Gamma(s+1) \zeta(2 s-2) ,
\end{equation*}
with the poles $s_i$ given above. Expression \eqref{eq:res_sum} is correct, if the integration along the contour around all the poles to the left of $s_0$ vanishes. It can be shown that this is indeed the case by carefully analyzing the asymptotic behavior of $\zeta(2 s-2)$ and $\Gamma(s \pm 1)$ (see e.g.\ Ref.~\cite{paris2018evaluation}).

Now, we do not have to consider all the residues in \eqref{eq:res_sum}, because we are only interested in the terms of $S(\tau)$ that are not higher than 2nd order in $\tau$, since all higher orders will lead to positive powers of $a$ in \eqref{eq:E_reg_5dim_large_rc}. But these will not contribute to the Casimir energy in the limit $a \to 0$. Hence, we only have to consider the poles higher than and up to $s=-1$. We find
\begin{equation}
S(\tau) = \hal \left[ \frac{3 \pi}{\tau^3} - \frac{2}{\tau^2} + \frac{\zeta'(-4)}{4} \tau^2 + \mac{O} \left( \tau^4 \right) \right] ,  \label{eq:res_sum_result}
\end{equation}
where $\zeta'$ is the derivative of $\zeta$. Inserting \eqref{eq:res_sum_result} in \eqref{eq:E_reg_5dim_large_rc}, we get
\begin{IEEEeqnarray}{rCl} 
E^{\te{reg}}(r_c \to \infty) &=& \frac{3}{2} \frac{\pi^2}{a^4} -\frac{\pi^2}{4} \left( - \frac{6 \pi}{a^5} d_1 + \frac{3}{a^4} + \frac{\zeta'(-4)}{8} \frac{1}{d_1^4} +\mac{O} \left( \frac{a^2}{d_1^2} \right) + (d_1 \rightarrow d_2)  \right) \nonumber \\
&=& \frac{3 \pi^3}{2} \frac{L}{a^5} - \frac{\pi^2 \zeta'(-4)}{32} \frac{1}{R^4} - \frac{\pi^2 \zeta'(-4)}{32} \frac{1}{(L-R)^4}   +\mac{O} \left( \frac{a^2}{d_i^2} \right) . 
\end{IEEEeqnarray}


\acknowledgments
 
We would like to thank Gia Dvali for encouraging the present investigation and giving valuable comments. Further we want to thank Marco Michel for his help with setting up the cluster computation and Simone Strohmair for her support with the fitting procedure.

\bibliographystyle{JHEP}
\bibliography{library_phd_thesis}

\end{document}